\documentclass[a4paper,11pt]{article}
\pdfoutput=1 

\usepackage{jheppub} 

\usepackage[T1]{fontenc} 

\usepackage{setspace}
\usepackage{amsmath}
\usepackage{tensor}
\usepackage{mathtools}
\usepackage{enumitem}
\usepackage{physics}
\usepackage{hyperref}
\usepackage{tcolorbox}
\usepackage{bbold}
\usepackage{tikz-cd}
\usepackage{rotating}
\usepackage{arydshln}
\usepackage{xcolor}
\usepackage{ulem}

\newcommand{\R}{\mathbb{R}}
\newcommand{\C}{\mathbb{C}}
\newcommand{\Z}{\mathbb{Z}}
\newcommand{\N}{\mathbb{N}}
\newcommand{\Q}{\mathbb{Q}}

\newcommand{\mF}{\mathcal{F}}

\newcommand{\id}{\mathbb{1}}
\newcommand{\M}{\mathcal{M}}

\newcommand{\mH}{\mathbb{H}}
\newcommand{\Gr}{\text{Gr}}

\newcommand{\mfe}{\mathfrak{e}}
\newcommand{\Arg}{\text{Arg}\,}
\newcommand{\Ln}{\text{Ln}\,}

\newcommand{\mM}{\mathcal{M}}
\newcommand{\NN}{\mathcal{N}}
\newcommand{\mS}{\mathcal{S}}

\newcommand{\SL}{\text{SL}}
\newcommand{\U}{\text{U}}
\newcommand{\SO}{\text{SO}}

\newcommand{\mO}{\text{O}}
\newcommand{\Mp}{\text{Mp}}

\newcommand{\SG}{\text{S}\Gamma}

\newcommand{\pl}{\partial}

\makeatletter
\newcommand{\Rom}[1]{\uppercase\expandafter{\romannumeral #1\relax}}
\newcommand{\rom}[1]{\lowercase\expandafter{\romannumeral #1\relax}}
\newcommand{\dg}{\dagger}
\newcommand{\overbar}[1]{\mkern 1.5mu\overline{\mkern-1.5mu#1\mkern-1.5mu}\mkern 1.5mu}

\def\longeq#1{
\begin{equation}
    \begin{aligned}
    #1
    \end{aligned}
\end{equation}
}

\def\lceq#1{
\begin{equation}
\begin{gathered}
    #1
\end{gathered}
\end{equation}
}

\def\mx#1{
\left(
\begin{matrix}
#1
\end{matrix}
\right)
}

\def\mathtitle#1#2{
\texorpdfstring{$#1$}{#2}
}

\DeclareUnicodeCharacter{0301}{\'{a}}
\DeclareUnicodeCharacter{0303}{\'{n}}

\newcommand{\by}{\Bar{y}}
\newcommand{\deno}{1 - 2 \Bar{y}_k y_k + y^2 \Bar{y}^2}

\newtheorem{thrm}{\color{blue} Theorem}\numberwithin{thrm}{section}
\newtheorem{defn}{\color{blue} Definition}\numberwithin{defn}{section}
\numberwithin{lem}{section}
\title{Consistency of eight-dimensional supergravities: \\
Anomalies, Lattices and Counterterms}

\author[a,b]{Bing-Xin Lao,}
\author[b]{Ruben Minasian}

\affiliation[a]{École Normale Supérieure - PSL, 45 rue d’Ulm, F-75230 Paris cedex 05, France}
\affiliation[b]{Institut de Physique Th\'{e}orique, Universit\'{e} Paris-Saclay, CNRS, CEA, F-9119, Gif-sur-Yvette,
France}

\emailAdd{bingxin.lao@ens.psl.eu}
\emailAdd{ruben.minasian@ipht.fr}

\abstract{We reexamine the question of quantum consistency of supergravities in eight dimensions. Theories with 16 supercharges suffer from the anomalies under the action of its discrete modular groups. In minimally supersymmetric theory coupled to Yang-Mills multiples of rank $l$ with the moduli space given by $\SO(2,l)/ (\U(1) \times \SO(l))$,   the existence of a counterterm together with the requirement that its poles and zeros correspond to the gauge symmetry enhancement imposes nontrivial constraints on the lattice. The counterterms needed for anomaly cancellation  for all cases, that are believed to lead to consistent theories of quantum gravity ($l = 2,10,18$), are discussed. }

\begin{document} 
\maketitle
\flushbottom

\section{Introduction and summary}\label{sec:Intro}
Existence of an anomaly cancellation mechanism in (super)gravity theories serves as a good guideline for selecting candidates for theories that can be consistent at the quantum level.

In minimally supersymmetric theories in ten dimensions, existence of Green-Schwarz mechanism reduces the number of possible choices for the gauge groups in the YM sector to four \cite{Taylor:2011wt}. From the other side, the existence of an anomaly inflow mechanism to two-dimensional chiral strings coupled to the theory restricts this number to two by ruling out the theories with abelian gauge factors \cite{
Adams:2010zy, Kim:2019vuc}. In six dimensions, there is an infinite number of anomaly-free minimal supergravities \cite{Taylor:2011wt}. Many, notably infinite families of $(1,0)$ theories, are ruled out by a closer examination of the inflow mechanism and the anomaly cancellation for two-dimensional $(0,4)$ strings coupled to the theory \cite{Kim:2019vuc, Lee:2019skh, Cheng:2021zjh}.

The focus of this paper is on eight-dimensional minimal supergravities. Classically, the 8D $\mathcal{N} = 1$ supergravity multiplet, made of a graviton, $B$-field, dilaton, two vector fields as well as spin-$\frac{3}{2}$  and spin-$\frac{1}{2}$ Majorana fermions (gravitino and a dilatino), can be coupled to any number of vector multiplets each comprising a vector field (photon), a gauginio (spin-$\frac{1}{2}$ Majorana fermion) and $2$ real scalars \cite{1985_PLB_Salam}. Supposing the number of vector multiplets is $l$, the $2l$ real scalars contained in the matter sector  parametrize the moduli space given by a K\"ahler manifold
\lceq{\label{eq:modsp}
\mM = \frac{\SO(2,l)}{\U(1) \times \SO(l)} \, .
}
These $l$ vectors together with two vectors in the gravity multiplet form an $(l+2)$-dimensional representation of $\SO(2,l)$.

The first restriction on admissible values of $l$ once more comes from anomalies - theories with odd numbers of Majorana fermions in 8D and 9D suffer from global anomalies \cite{Alvarez-Gaume:1983ihn}, and hence $l$ has to be even \cite{Montero2021}. There are further restrictions: 
\begin{itemize}
\item In theories with 16 supercharges in $D$ dimensions the number of vector multiples consistently coupled to gravity is bound by $26-D$ in order to assure the unitarity of strings couples to the theory \cite{Kim:2019ths}. Hence $l \leq 18$.
\vspace{-0,2cm}
\item Considering 8D theories on particular backgrounds and using 6D anomaly cancellation it has been argued that in fact the only admissible values of $l$ are $l=2$, $l=10$ and $l=18$ \cite{ Montero2021}.
\vspace{-0.2cm}
\item The symmetry enhancement (as well as the rank of of the YM algebra coupled to string probes) as predicted by the consistency of the supergravity  \cite{Hamada2021} is an agreement with the landscape of 8D string constructions \cite{Font:2020rsk, Font:2021uyw}. 
\item In the formulation of the theory with a four-form potential in the gravity multiplet, constraints on the global structure of the gauge groups can be deduced from the  the absence of anomalies between large gauge transformations of $B_4$ and 1-form symmetries  \cite{Cvetic:2020kuw, Cvetic:2021sjm}\footnote{Somewhat orthogonal to our discussion, global anomalies and topological analogues of Green-Schwarz mechanism have been discussed in 8D with 16 supercharges \cite{Garcia-Etxebarria:2017crf} and in 10D type IIB theory 
\cite{Debray:2021vob}. In this paper we are mostly concerned with the existence of local counterterms.}.
\end{itemize}
\noindent
We would like to reexamine these results from the point of view of 8D anomaly cancellation. Neither the  $\mathcal{N} = 1$ theory nor its $\mathcal{N} = 2$ counterpart, where the scalars parametrize the $\SL(2)\times \SL(3)/\left(\U(1)\times \SO(3) \right)$ coset, suffers from chiral anomalies. However, both theories, with 16 and 32 supercharges, have local anomalies under the composite $\U(1)$ in the denominator of the coset.

The moduli space of  supergravity theories with extended supersymmetry typically has scalars parametrizing a coset $G/H$. The numerator of the coset, $G$, denotes the U-duality group of the theory, and some discrete version of it gives rise to an exact symmetry after quantization. Theory can be formulated in a way that $G$ acts only on bosonic fields. The denominator $H$, which is the maximal compact subgroup of $G$, is regarded as a gauge symmetry of the theory. Indeed, the compact part of the Cartan-Maurer form of the coset element transforms as a gauge field under the $H$ transformations. The supersymmetry variations of all fermionic fields, which are inert under $G$, involve this composite connection corresponding to $H$. When $H$ contains a $\U(1)$ factor, it may couple to fermions in a chiral fashion, a priori giving rise to a composite chiral anomaly \cite{Marcus1985}. This is exactly what happens in eight dimensions.

The physical content of the theory is usually identified by fixing the gauge, thereby eliminating the redundant bosonic degrees of freedom associated to $H$. When the local symmetry is gauge fixed, the U-duality becomes non-linearly realized. Moreover, the fermionic fields now transform under $G$. Part of this transformation may still be realized as a nontrivial phase shift. Therefore, the gauge fixing translates the $\U(1)$ anomaly into an anomaly under the surviving discrete part of $G$, making the theory ill-defined.

The existence of this anomaly implies that the  symmetry group $\SL(2;\R)$ may not be continuously maintained in the quantum theory. For the theory to be consistent, a cancellation mechanism should be figured out, in the process deciding to what extent the symmetry survives. The question is if it can be done by the addition of a local counterterm with appropriate modular properties under the transformation of the discrete version of $G$. Originally such counterterm was discussed in the context of ten-dimensional IIB string theory \cite{Gaberdiel_1998}, but the formalism is adapted to 8D theories as well \cite{Gaberdiel_1998, Minasian2017}. An (in)ability of finding such a counterterm is the reason why the value of $l$ and the lattice structure of the gauge group in 8D get restricted.\footnote{The construction of the counterterms naturally introduces modular forms. A review of modular forms and their important applications in physics can be found in \cite{Hoker2022}.} 

Let us outline the anomaly cancellation mechanism. Denoting the anomalous composite $\U(1)$ connection by $Q$ and its curvature by $F^Q$, the anomaly is given by the descent formula from the ten-form anomaly polynomial 
\lceq{\label{eq:anogen} I_{10} = \frac{F^Q}{2\pi} \wedge X_8 \, ,}
where $X_8=X_8(R, \mF)$ is a polynomial in $R$ and the non-abelian gauge field strength $\mF$ for $\mathcal{N} = 1$ case (the exact form of  the polynomial in this case on the contrary is not going to play any role in our discussion). The resulting anomalous phase variation in the partition function $\Delta = - \int \Sigma \, X_8$ can locally be cancelled by adding a term to the action 
\lceq{ \mathcal{S}_{\phi} = \int  \phi\, X_8 \, ,}
where $\phi$ is a scalar degree of freedom transforming under $\U(1)$: $\phi \rightarrow \phi + \Sigma$. This $\phi$ can be set to zero by gauge fixing (think of the third scalar in $\SL(2;\R)$), but since $\delta_M \phi \neq 0$ under the $G$-valued transformation $M$, the local counterterm is not $G$-invariant. In other words, the local $\U(1)$ anomaly transfers to the global $\SL(2;\R)$ anomaly. In principle such a breaking of duality symmetry does not introduce inconsistency into the theory. However, extra considerations are needed if we want to maintain this duality group to some extent. As shown explicitly for $\SL(2; \R)$ in \cite{Gaberdiel_1998} and will be extended to $\SO(2,l;\R)$ here, one can design a counterterm $S$ such that under the $G$-valued transformation $M$
\lceq{\label{eq:cancel} \delta_M \mathcal{S} = - \delta_M \mathcal{S}_{\phi}   +  \arg \chi(M) \int   X_8\, ,}
where $\chi(M)$ is a phase factor and $\delta_M \mathcal{S}_{\phi} =  \int \Sigma \, X_8$. If this phase factor $\chi(M) \equiv 1$ for any $M \in G$ there will be a complete anomaly cancellation but that does not always happen. Note that in general it should suffice that the partition function is well-defined, and hence $\delta_M$ of the entire action integrates to an integer (times $2 \pi$). The value of $\chi(M)$ depends on $l$ and on the details of the lattice of signature $(2,l)$, which will naturally appear during the construction of counterterms. So at the first glance this presents a dilemma: either one should be imposing case-by-case integrality conditions on $\int X_8 (R, \mF)$ or, as we shall argue, opt for a universal consistency condition and require that $\chi(M) = 1$ for every  $\mathcal{N} = 1$ theory. Regardless of philosophy, let us turn to the details of how \eqref{eq:cancel} works. The first important point is the precise form of $\delta_M \phi$.  For instance, in 10D Type \Rom{2}B supergravity or in $\NN=2$ theory in 8D,  the coset element of $\SL(2)/\U(1)$  is parametrized by the modular parameter $\tau$ and the compensating $\U(1)$ transformation under the $\SL(2;\R)$  takes the  form
\lceq{\label{eq:sec1_U(1)trans}
e^{- i \Sigma (M, \tau)} = \left(\frac{c \tau + d}{c \bar{\tau} + d}\right)^{\frac{1}{2}} \, , \quad M \in \mx{
a & b \\
c & d
}, \quad M \in \SL(2;\R)\,, \quad \tau \in \mathbb{H} \, .
}
The second crucial  point is that there exists a function of $\tau$, the Dedekind eta function $\eta(\tau)$, that under $\SL(2; \Z)$ transformation picks a factor $\sim (c\tau +d)^{1/2}$. As a consequence, a ratio of $\eta(\tau)$ and its complex conjugate can be used in constructing the counterterm \cite{Gaberdiel_1998}. 

In this paper we are mainly interested in the eight-dimensional supergravity with $16$ supercharges, and the moduli space of the theory is given by  $\mathcal{M} $ \eqref{eq:modsp}. The moduli space $\mM$ is a realization of the hermitian symmetric space. Moreover, the tube domain,  called the generalized upper-half plane $\mH_l$, can be realized in this space  \cite{bruinier2002borcherds, bruinier20081}. We find that the generalized upper-half plane $\mH_l$ provides the correct framework to describe the gauge transformations. By introducing Calabi-Vesentini coordinates \cite{Calabi_1960_AM}, we explicitly compute the $\U(1)$ gauge potential and its field strength, and show how the $\U(1)$ compensating transformation  generalizes equation~(\ref{eq:sec1_U(1)trans}).  It is formed by the so called automorphy factor $j(M,Z)$ (where $M$ is an $\SO(2,l; \R)$ transformation, and $Z$ denotes the coordinates on the generalized upper half plane):
  \lceq{\label{eq:compensatingU(1)}
 e^{-i \Sigma(M,Z)} = \frac{j(M,Z)}{|j(M,Z)|} \, .
}
Equivalently we have $- \Sigma = \arg j(M,Z) = \Arg j(M,Z) + 2 k \pi$ for $k \in \Z$ and $\Arg$ denotes the principal branch of the argument taking the value from $[-\pi,\pi)$. Finding a  function $\Psi(Z)$ such that 
\lceq{ \label{eq:modular_form}
\Psi (M \langle Z \rangle ) = \chi (M) j(M, Z)^r \Psi(Z)
}
would allow to construct the counterterm $\mS$ as 
\lceq{\label{eq:countert}
\mathcal{S} = \frac{1}{r} \int \arg \Psi(Z) X_8 \, .
}
Indeed such functions, or more precisely  meromorphic modular forms on the orthogonal group $\mO(2,l)$  of weight $r$ and multiplier system $\chi$, can be found by using the Borcherds products \cite{Borcherds1995}.
The original discovery that the automorphic forms on $\mO^+(2,s+2)$ ($l = s + 2$) can be written as infinite products was made in the context of unimodular latices. Following the use of theta correspondence, which gave an alternative approach to these results  \cite{HARVEY1996315}, the generalisation of the constructions of modular forms to non-unimodular lattices was provided \cite{Borcherds1998}. The case $l = 2$ case requires special treatment because, strictly speaking, the Borcherds product does not apply and an alternative derivation of the counterterms is needed, which we left for our future work.

As we shall see, the requirement that the modular form $\Psi(Z)$ that exists, ensuring the complete anomaly cancellation, is not particularly restrictive. However there is an additional consideration: the local counterterms constructed from the meromorphic $\Psi(Z)$ are not well defined at its zeros or poles. For Borcherds products, these points lie on the so-called rational quadratic divisors (RQD). In fact some of these divisors have physical interpretation and correspond to the symmetry enhancement points in the moduli space~\cite{HARVEY1996315}.\footnote{Note that this observation was first made in the context of threshold corrections 4D $\NN=2$ theories, and the functions involved are required to be automorphic. Here we need anomaly cancelling counterterms, which require modular forms of non-trivial weights. We shall return to the comparison of the 4D and 8D in section \ref{sec:conclusion}.} For these, the theory will continue being consistent even if the counterterm is not well-defined. Moreover the gauge symmetry enhancement should be in agreement with the symmetries of the lattice. We will  show that these physical constrains  lead to the requirement that the lattice is {\sl reflective} (defined in  section~\ref{sec:general_strategy}). The number of reflective lattices is finite and their rank is bounded by $l=26$. These bounds are less stringent than those imposed by swampland.

The organisation of this paper is as follows. In section~\ref{sec:Mathematical_preliminaries} we spell out the anomaly that needs to be cancelled, and  introduce the necessary mathematical preliminaries needed for the construction of the counterterms (with further details collected in appendix~\ref{app:supplement_modular_form}). Section~\ref{sec:U(1)_anomaly_in_8D_SUGRA_general} is devoted to  the derivation of the compensating $\U(1)$ transformation (equation~\ref{eq:compensatingU(1)}). The construction of counterterms is presented in  section~\ref{sec:general_strategy}. In this section we also consider the implications  of zeros and poles of the modular forms and the ensuing constraints on admissible lattices. The discussion mainly focuses on the case $l \ge 3$ while we make some comments on $l = 2 $ case at the end. A brief summary and discussion of some open questions are presented in section~\ref{sec:conclusion}.

\section{Minimal supergravity and lattices of\mathtitle{(2,l)}{(2,l)}signature} \label{sec:Mathematical_preliminaries}
Generically the minimal supergravity in 8D comprises a single gravity multiplet and $l$ vector multiplets. The field content of $\mathcal{N}=1, D=8$ supergravity is  given by
\lceq{
\left(\tensor{e}{_\mu^m}, \psi_\mu, \chi, B_{\mu \nu}, \tensor{A}{_\mu^i}, \sigma \right)\, , \quad i = 1,2,
}
where $\tensor{e}{_\mu^m}$ is the graviton, $\psi_\mu$ is the gravitino, $\chi$ is dilatino, $B_{\mu \nu}$ is the antisymmetric tensor (background field), $\sigma$ is the dilaton. Both $\psi_\mu$ and $\chi$ are pseudo-Majorana spinors. $\tensor{A}{_\mu^i}$  and the scalar $\sigma$ are real. Coupling $l$ vector multiplets of the form $(\lambda, A_\mu, \phi^i)$ and combining the field content together we obtain
\lceq{\label{eq:minsugra}
\left(\tensor{e}{_\mu^m}, \psi_\mu, \chi, B_{\mu \nu}, \tensor{A}{_\mu^I}, \phi^\alpha, \sigma\right),
}
where $I=0, \ldots, l+1$, $\alpha = 1,\ldots, 2l$. Here we adopt the metric convention 
\lceq{
\eta_{AB} = \eta_{IJ} = \left(+1,+1,-1,\ldots, -1\right) \, .
}
The $2l$ real scalars $\phi^\alpha$ parameterize the moduli space
\lceq{
\mathcal{M} = \frac{\SO(2,l)}{\SO(2)\times \SO(l)} \cong \frac{\SO(2,l)}{\U(1)\times \SO(l)} \, .
}
The fermions of the theory have chiral couplings to one of the composite $\U(1)$ in the denominators of the coset \cite{1985_PLB_Salam}. The $\U(1)$ charges of the gravitino (positive chirality), the dilatino (negative chirality) and the gaugini (positive chirality) are all $\frac{1}{2}$.\footnote{Notice that the same charge assignment is valid in dual formulation of the theory where the two-form $B$ is replaced by a four-from \cite{Awada:1985ag}, and the discussion of the counterterms applies to both.} Hence, the anomaly polynomial is 
\lceq{
I_{8D} = I_{3/2} - I_{1/2}^{\text{dilatino}} + I_{1/2}^{\text{gaugini}} \, .
}
If the gauge group is given by $G$ ($\text{rank}(G) = l$), the gaugini couple both to $G$ (the fields strength of the gauge field will be denoted by $\mF$) and the composite $\U(1)$ (whose field strength is again denoted by $F^Q=d Q$). The resulting polynomial is of the form \eqref{eq:anogen} with  (see \cite{Minasian2017} for details)
\longeq{\label{eq:anomalypol}
& X_8  (R,  \mF) =
\\
& \frac{1}{32 (2 \pi)^3} \left[(248 + \dim G) \left[\frac{\tr R^4}{360} + \frac{(\tr R^2)^2}{288}\right] - (\tr R^2)^2 + \frac{1}{6}\tr R^2 \Tr \mathcal{F}^2 + \frac{2}{3} \Tr \mathcal{F}^4\right] \, ,
}
and given the variation $\delta Q = d \Sigma$ the anomalous phase is 
\longeq{\label{eq:sec3_anomalous_phase}
\Delta_G = - \int \Sigma \, X_8(R, \mF)  \, ,
}
The precise form of $X_8(R, \mF)$ in \eqref{eq:sec3_anomalous_phase} is not important for our discussion. The idea is to constrain the admissible theories and their lattices rather than try to cancel \eqref{eq:sec3_anomalous_phase} by imposing case by case conditions on the integrality  properties of $X_8(R, \mF)$. In addition, the counterterm can be changed by adding massive states and integrating them out. While the role of the massive completions presents interesting questions, here we are concerned by the possibility of writing a counterterm that will lead to an anomaly cancellation for any (discrete) $\SO(2,l)$ transformation.

As we introduce in the introduction, we will have to construct the compensating $\U(1)$ transformation with respect to this group (see Sec. \ref{sec:U(1)_anomaly_in_8D_SUGRA_general}) and find the modular forms with right properties to serve as counterterms \eqref{eq:countert}. 

In the rest of this section we will discuss some of the necessary background and set up the notation. In order to make the presentation self-contained, we will include some of the basic definitions. Further details can be found in Appendix \ref{app:supplement_modular_form}, where  the presentation follows closely to the references \cite{bruinier2002borcherds, bruinier20081}.

\subsection{Lattices of\mathtitle{(2,l)}{(2,l)}signature and generalized upper half plane}
A typical lattice $L$ in $\R^b$ has the form $L := \left\{\sum_{i=1}^b a_i v_i | a_i \in \Z\right\}$, where $\{v_1, \ldots v_b\}$ is the  basis set. Usually the lattice is equipped with a quadratic form $q: L \rightarrow \R$, which defines the norm of the vector $x$ in the lattice as $q(x)$ and  naturally induces a symmetric bilinear form $( \cdot , \cdot ): \Lambda \times \Lambda \rightarrow \R$
\lceq{
( x , y ) :=  q(x+y) - q(x) - q(y)\, , \quad \text{for} \quad x,y \in L \, .
}
It is easy to see that $q(x) = \frac{1}{2} ( x, x )$ since $q$ is a quadratic bilinear form. The lattice is called even if $q(x) \in \Z$  for arbitrary $x \in L$. The dual lattice $L'$ is defined as 
\lceq{
L' := \left\{y\in L \otimes \Q|\, (  y, x ) \in \Z \,  \, \text{for} \, \,  \forall x \in L \right\} \, .
}
A lattice $L$ is called self-dual or unimodular if it is equal to its dual $L = L'$. The quadratic form $q$ has signature, denoted by $(b^+,b^{-})$ and $b^{+}+b^{-} = b$, where $b^+$ ($b^-$) denotes the number of the $+$ ($-$) signs. An important theorem states that there are no indefinite even unimodular lattices unless $ b^+ - b^- \equiv 0 \mod 8$. 

Let us denote the lattice and its quadratic form by a pair $(L, q)$.
Suppose $\left(L,q\right)$ is a lattice that has a signature $(2,l)$. Consider the Grassmannian of 2-dimensional subspaces of $V = L \otimes \R$ on which the quadratic form is positive definite 
\lceq{\label{eq:sec4_Grassmannian}
\Gr_2 (V) := \left\{v \subset V | \dim v = 2 \, \text{and} \, q|_{v} > 0\right\} \, ,
}
where $q|_{v}>0$ means that for every element $x \in v$, $q(x)>0$\footnote{We have defined $q$ on the lattice $L$, i.e. $q(v_i)$ has a clear definition for $1\le i \le l+2$. With the help of the induced bilinear form $( x, y ) = q(x+y) - q(x) - q(y)$, we can safely extend the quadratic form to the space $V= L \otimes \R$. }. We define the orthogonal group and the special orthogonal group as 
\lceq{ \notag
\mO \left(V;\R\right) := \left\{ \sigma \in \text{Aut}(V) | \,  \sigma \, \text{is an isometry of $V$}\right\}, \, \,  
\SO \left(V;\R\right) := \left\{\sigma \in \mO(V;\R)| \det \sigma = 1\right\} \, .
}
Since $V$ is a usual linear space on $\R^{l+2}$, one can think of these two as matrix groups. If two spaces $V_1, V_2$ have the same signature, it can be proved that the orthogonal groups are isomorphic, i.e. $\mO(V_1; \R) \cong \mO(V_2; \R)$. Thus we can denote the (special) orthogonal group by using the signature like $\mO(2,l;\R)$ ($\SO(2,l ;\R)$). One can prove that $\mO(2,l;\R)$ acts transitively on $\Gr_2(V)$. If $v_0 \in \Gr_2(V)$ is fixed, the stabilizer $K$ of $v_0$ is a maximal compact subgroup of $\mO(2,l;\R)$ and $K\cong \mO(2)\times \mO(l)$. This  constructs an isomorphism $\Gr_2(V) \cong \mO(2,l;\R)/K$, which is a realization of the hermitian symmetric space. 

To see the complex structure, we consider the complexification $V(\C) = V \otimes \C$ of $V$. Since $V$ is has negative signatures, there exists non-trivial isotropic vector $x$, which satisfies $q(x) = 0$ and $x \neq 0$. The isotropic subspace (or called the zero quadric) is 
\lceq{
\mathcal{I} := \left\{ Z_L \in V(\C)\backslash \{0\}| \left( Z_L ,Z_L  \right) = 0\right\}\, .
}
We consider the projective space $P\mathcal{I} := \mathcal{I}/\sim $, where the equivalence relation is $Z_L \sim t Z_L$ for arbitrary $ t \in \C, \, t \neq 0$. The equivalence class can be denoted as $[Z_L]$. Consider the subset 
\lceq{\label{eq:sec4_mathcalK}
\mathcal{K} := \left\{ [Z_L] \in P\mathcal{I}| \, ( Z_L,Z_L ) = 0, \, ( Z_L, \overbar{Z_L} ) > 0\right\} \, ,
}
$\mathcal{K}$ is a complex manifold of dimension $l$ consisting of two connected components. The subgroup $\mO^+(2,l;\R)$ of elements whose spinor norm equals the determinant preserves the components of $\mathcal{K}$, whereas $\mO(2,l;\R)\backslash \mO^+(2,l;\R)$ interchanges them. We can denote the components $\mathcal{K}^+$ and $\mathcal{K}^-$ respectively.

For arbitrary $Z_L \in V(\C)$, we can write $Z_L = X_L + i Y_L$, $X_L, Y_L \in V$, and construct a map between $\mathcal{K}^+$ and $\Gr_2(V)$
\lceq{\label{eq:sec4_complex_structure_Gr}
[Z_L] \longmapsto v(Z_L) =  \left\{a X_L + bY_L | \,  a, b \in \R\right\} \, .
}
This map is an analytic isomorphism.  Thus, the set $\mathcal{K}^+$ and the map indeed provide a complex structure on $\Gr_2(V)$. 
We refer the reader to \cite{bruinier20081} for detailed proof.

An important concept, useful for us,  is that of the generalized upper-half plane corresponding to the $\SO(2,l;\R)$ transformation.  We will first describe the formal way of constructing the generalized upper-half plane $\mH_l$~\cite{bruinier20081}. A  specific method for achieving the generalization \cite{Williams2021, krieg2016integral}, which is more appropriate for our discussion, will be discussed later.

Suppose $z \in L$ is a primitive norm zero vector, i.e. $q(z) = 0$ and $\Q z \cap L = \Z z $. Let $z' \in L'$ be another vector which satisfies $(z,z') = 1$. We define the sub-lattice $K$
\lceq{
K := L \cap z^\perp \cap z'^\perp \, ,
}
where $z^\perp$ denotes the orthogonal subspace of $z$, such that that all vectors $x$ in this subspace satisfy $( x, z) = 0$. Then $K$ is Lorentzian, i.e. of the signature is $(1,l-1)$. The space $V$ can be decomposed into 
\lceq{
V =  \R z \oplus \left(K \otimes \R\right) \oplus \R z' \, , \quad V(\C) =  \C z \oplus \left(K \otimes \C\right) \oplus \C z' \, .
}
For arbitrary vector $Z_L \in V(\C)$, there exists a unique combination $\left(a, Z, b\right)$ such that $Z_L = a z + Z + b z'$, $Z \in K \otimes \C$, $a,b \in \C$, which means that we can use the combination $(a,Z, b)$ to represent a vector. We define a set $\widetilde{\mH}_l$
\lceq{\label{eq:sec4_tildeHn}
\widetilde{\mH}_l = \left\{Z = X + i Y \in K \otimes \C|\, X, Y  \in K\otimes \R, \,  q(Y) > 0 \right\}  \, .
}
Since the lattice $K$ has signature $(1,l-1)$, the set of positive-norm vectors in $K \otimes \R$ splits into two connected components, $\mathcal{K}^{\pm}$. We define the map $f$ 
\longeq{\label{eq:sec4_decomposeZ}
f: \widetilde{\mH}_n &\longrightarrow \mathcal{K} \\ 
Z &\longmapsto f(Z) = \left[\left(-q(Z) - q(z'), Z, 1\right)\right] \, ,
}
where $\left(-q(Z) - q(z'), Z, 1\right)$ is an $l+2$ dimensional vector is space $V(\C)$. As the references~\cite{bruinier20081,bruinier2002borcherds} show, $f$ is a biholomorphic map. Under the map $f$, two connected components of $\widetilde{\mH}_l$ map into the two connected components $\mathcal{K}^{\pm}$ of $\mathcal{K}$ separately. We choose $\mH_l$ to be the component of $\widetilde{\mH}_l$ that maps  $\mathcal{K}^+$. This realization of $\mathcal{K}^+$ as a tube domain can be viewed as generalized upper-half plane $\R^l + i \Omega^l$, where $\Omega^l$ is the positive-norm cone. 

\subsection{Action of\mathtitle{\mO(2,l)}{O(2,l)}on generalized upper-half plane}\label{subsec:sec4_action_O_on_generalized_plane}
To construct the generalized upper-half we will split the lattice $(L, q)$ into $(L,q) = (L_0, q_0) \oplus \Pi_{1,1}$ where $(L_0, q_0)$ is a lattice of signature $(1,l-1)$, equipped with the quadratic form $q_0$ and $\Pi_{1,1}$ is the unimodular lattice with signature $(1,1)$ equipped with the quadratic form $q((a,b)) = ab$. The generalized upper-half plane can be defined in the following way
\lceq{\label{eq:sec4_Hn}
\mH_l = \left\{Z = X + i Y \in L_0 \otimes \C |X,Y \in L_0 \otimes \R, Y \in P\right\} \, ,
}
where $P$ denotes the future light cone of the Minkowski space $L_0 \otimes \R$.

We set $l = s + 2$ and take $s \ge 0$ and even throughout the discussion. Suppose $\hat{S}$ is an $s \times s $ symmetric positive definite real matrix (when $s = 0$, $\hat{S}$ collapses, but the discussion bellow still applies) and define 
\lceq{\label{eq:sec4_qudratic_form}
S_0 = \mx{
& & 1 \\
&  -\hat{S} & \\
1 & & 
} \in \text{Sym} \left(s+2 ;  \R\right) \, ,  \\
S = \mx{
& & 1 \\
& S_0 & \\
1 & &  
} \in \text{Sym} \left(s+4 ;  \R\right) \, ,
}
where $\text{Sym} (s+2 ; \R)$ denotes the set of $(s+2) \times (s+2)$ symmetric real matrix. We then define for $Z_L \in L$ (or $L \otimes \C$), $q(Z_L) = \frac{1}{2} Z_L^T S Z_L$  and for $Z \in L_0$ (or $L_0 \otimes \C$), $q_0 (Z) = \frac{1}{2} Z^T S_0 Z$. Here the superscript $T$ means the transpose operation. 


In the following  we will frequently use the notation $L(N)$ for a positive integer $N$ (Some references use the notation $\sqrt{N}L$). $L(N)$ indicates the lattice with the same basis as $L$ but equipped with the scaled quadratic form $N S$ (or vectors scaled by factor $\sqrt{N}$ equivalently). We  consider explicitly only the lattices $L = \Pi_{1,1} (1) \oplus \Pi_{1,1}(1) \oplus \hat{L}$ with lattice $\hat{L}$ equipped with the quadratic form $\hat{S}$. The discussion can be extended to $L = \Pi_{1,1}(N_1) \oplus \Pi_{1,1} (N_2) \oplus \hat{L}$ for arbitrary positive integer $N_{1,2}$ and the results still apply.

With the help of these quadratic forms, by setting $z = \left(1, 0,\ldots , 0\right)^T \in L$ and $z' = \left(0,\ldots, 0,1\right)^T \in L'$, the map $f$ \eqref{eq:sec4_decomposeZ} can be written as 
\longeq{\label{eq:sec4_map_Hl_K+}
f: \mH_l & \longrightarrow \mathcal{K}^+ \\
   Z & \longmapsto [Z_L] =  \left[\left(-q_0(Z) , Z , 1\right)^T\right]  \, .
}
The (special) orthogonal transformation $\mO(2,l;\R)$ ($\SO(2,l;\R)$)\footnote{The definition here, while using a different metric,  is isomorphic to the definition using  $\eta = (+1, +1, -1, \ldots, -1)$.} now is 
\lceq{
\mO(2,l;\R) = \left\{M \in \text{Mat}(l+4; \R)| M^T S M = S\right\} \, ,\\
\SO(2,l;\R) = \left\{M \in \text{SL}(l+4; \R)| M^T S M = S\right\} \, .
}
Since $\mathcal{K}^+$ and $\mH_l$ are isomorphic, the action of $\mO^+(2,l;\R)$ on the set $\mathcal{K}^+$ naturally induces the action on the generalized upper-half plane $\mH_l$. If $M \in \mO^+(2,l;\R)$, the action on $\mathcal{K}^+$ is defined as
\longeq{
M: \mathcal{K}^+ &\longrightarrow \mathcal{K}^+ \\
 [Z_L] &\longmapsto [M Z_L] \, ,
}
where $M Z_L$ is the usual linear transformation of a vector in $\C^{l+2}$. Because the real orthogonal transformation will not change the the norm $( Z_L, Z_L ) $ and $( Z_L, \overbar{Z_L} ) $, the element $[M Z_L]$ still stays in the set $\mathcal{K}^+$. Then we can define the action of $M$ on the generalized upper-half plane $\mH_l$, $Z \mapsto M \langle Z \rangle  $, such that the following diagram commutes
\begin{equation}
\tikzset{every picture/.style={line width=0.75pt}} 
\begin{tikzpicture}[x=0.75pt,y=0.75pt,yscale=-1,xscale=1, baseline=(current bounding box.center)]
\draw    (70,48.5) -- (188,48.5) ;
\draw [shift={(190,48.5)}, rotate = 180] [color={rgb, 255:red, 0; green, 0; blue, 0 }  ][line width=0.75]    (10.93,-3.29) .. controls (6.95,-1.4) and (3.31,-0.3) .. (0,0) .. controls (3.31,0.3) and (6.95,1.4) .. (10.93,3.29)   ;
\draw    (70,148.5) -- (188,148.5) ;
\draw [shift={(190,148.5)}, rotate = 180] [color={rgb, 255:red, 0; green, 0; blue, 0 }  ][line width=0.75]    (10.93,-3.29) .. controls (6.95,-1.4) and (3.31,-0.3) .. (0,0) .. controls (3.31,0.3) and (6.95,1.4) .. (10.93,3.29)   ;
\draw    (50,128.5) -- (50,70.5) ;
\draw [shift={(50,68.5)}, rotate = 90] [color={rgb, 255:red, 0; green, 0; blue, 0 }  ][line width=0.75]    (10.93,-3.29) .. controls (6.95,-1.4) and (3.31,-0.3) .. (0,0) .. controls (3.31,0.3) and (6.95,1.4) .. (10.93,3.29)   ;
\draw    (201,128.5) -- (201,70.5) ;
\draw [shift={(201,68.5)}, rotate = 90] [color={rgb, 255:red, 0; green, 0; blue, 0 }  ][line width=0.75]    (10.93,-3.29) .. controls (6.95,-1.4) and (3.31,-0.3) .. (0,0) .. controls (3.31,0.3) and (6.95,1.4) .. (10.93,3.29)   ;
\draw (68,48.5) node [anchor=east] [inner sep=0.75pt]   [align=left] {$\displaystyle \mathcal{K}^{+}$};
\draw (192,48.5) node [anchor=west] [inner sep=0.75pt]   [align=left] {$\displaystyle \mathcal{K}^{+}$};
\draw (130,45.5) node [anchor=south] [inner sep=0.75pt]   [align=left] {$\displaystyle [ Z_L] \mapsto [ M Z_L]$};
\draw (68,148.5) node [anchor=east] [inner sep=0.75pt]   [align=left] {$\displaystyle \mathbb{H}_{l}$};
\draw (192,148.5) node [anchor=west] [inner sep=0.75pt]   [align=left] {$\displaystyle \mathbb{H}_{l}$};
\draw (48,98.5) node [anchor=east] [inner sep=0.75pt]   [align=left] {$\displaystyle f$};
\draw (130,145.5) node [anchor=south] [inner sep=0.75pt]   [align=left] {$\displaystyle Z\mapsto M \langle Z \rangle $};
\draw (203,98.5) node [anchor=west] [inner sep=0.75pt]   [align=left] {$\displaystyle f$};
\end{tikzpicture} \, . 
\end{equation}
In the equation form we have
\lceq{\label{eq:sec4_action_M}
\left[M f(Z)\right] = \left[f \left(M\langle Z \rangle \right)\right] \, .
}
For convenience, we decompose the matrix $M$ in the following way
\lceq{ \label{eq:sec4_action_M2}
M = \mx{
\alpha & a^T & \beta \\
b & P & c \\
\gamma & d^T & \delta 
} \in \mO^+(2,l;\R) \, , \quad  
\begin{cases}
\alpha, \beta , \gamma, \delta \in \R, \\
a,b,c,d \in \R^{l}, \\
P \in \text{Mat}\left(l;\R\right) \, .
\end{cases}
}
Expanding the equation~(\ref{eq:sec4_action_M}), we have 
\lceq{\label{eq:sec4_action_M3}
\mx{
\alpha & a^T & \beta \\
b & P & c \\
\gamma & d^T & \delta 
} \mx{
-q_0(Z) \\
Z \\
1
} = \mx{
-\alpha q_0(Z) + a^T Z + \beta \\
- b q_0(Z) + P Z + c \\
-\gamma q_0(Z) + d^T Z+ \delta 
} = j(M,Z) \mx{
-q_0(W) \\
W \\
1
} \, ,
}
where $W = M \langle Z \rangle$, $j(M,Z) \in \C$. From this equation we extract the definition the action of (special) orthogonal group on the generalized upper-half plane $\mH_l$ directly
\longeq{\label{eq:sec4_action_M4}
&W = M\langle Z \rangle  := \left(-b q_0 (Z) + P Z + c\right) \left(-\gamma q_0 (Z) + d^T Z + \delta\right)^{-1} \, ,\\
&j(M, Z) := -\gamma q_0(Z) + d^T Z + \delta  \, .
}
With such definition, the equality (\ref{eq:sec4_action_M3}) clearly holds for last two components. 
The first components of the vector on two sides  of (\ref{eq:sec4_action_M3}) are equal due to the norm-zero condition.

\subsection{Modular forms on generalized upper-half plane}
\label{sub:orthogonal_modular_forms}
As proposed in the section~\ref{sec:Intro}, a specific function transforming in the particular way under the modular group~(\ref{eq:modular_form}) will play a central role in the construction of the counterterm. We turn now to the modular forms on generalized upper-half plane, commonly referred to as orthogonal modular forms. 
We will restrict for now to  $l \geq 3$, where the application of the results of Borcherds \cite{Borcherds1995, Borcherds1998} apply.

The pivotal observation is that the group $\SL(2;\R)$ and $\mO(2,l;\R)$ form a dual reductive pair. To construct the modular forms of orthogonal group $\mO(2,l)$,   the modular forms on $\SL(2,\Z)$ can be lifted by integrating against the Siegel theta function $\Theta(\tau, Z)$ \cite{HARVEY1996315, Borcherds1998}. A brief review of the relevant background, following the presentation of~\cite{bruinier2002borcherds} is given  in appendix~\ref{app:supplement_modular_form}.

Suppose $\mO(L)$ is the orthogonal group of a even lattice $L$ with signature $(2,l)$ defined by
\lceq{
\mO(L) := \left\{M \in \mO(2,l;\R)| \, M L = L \right\} \, .
}
The orthogonal group of the discriminant group $D(L):= L'/L$ can be defined similarly and will be denoted as $\mO(L'/L)$. We then denote by $\mO_d(L)$ the discriminant kernel of $\mO(L)$, which is the subgroup of finite index of $\mO(L)$ consisting of all elements which act trivially on the discriminant group $L' /L$, i.e.
\lceq{\label{eq:sec4_OdL}
\mO_d (L) := \text{Ker} \left(\mO(L) \rightarrow \mO(L' / L)\right) \, .
}
We define the intersection with $\mO^+ (V)$, $V = L \otimes \R$ as the modular group
\lceq{\label{eq:sec4_GammaL}
\Gamma(L) := \mO^+ (V) \cap \mO_d (L) \, .
}
Recalling the definition of $j(M, Z)$, we can rewrite it as $
j(M, Z) = \left(M Z_L, z\right)$ with $l+2$-dimensional vectors $z = \left(1,0,\ldots, 0\right)^T$ and $Z_L = (- q_0 (Z), Z, 1)^T$. Following   Theorem 13.3 in \cite{Borcherds1998} (Theorem~\ref{Thrm:sec4_Borcherds}), we can lift a nearly holomorphic modular form $f(\tau) = \sum_{\gamma \in L'/L} f_\gamma \mfe_\gamma: \mH \rightarrow \C[L'/L]$ (see Definition~\ref{defn:appB_nhmf}) of weight $1 - l/2$ with Fourier expansion
\lceq{\label{sec3:eq_nhmf}
f(\tau) = \sum_{\gamma \in L'/L} \sum_{n \in \Z + q(\gamma)} c(\gamma,n) \mfe_\gamma(n \tau) \, ,
}
to the meromorphic function $\Psi(Z) : \mH_l \rightarrow \C$ with the following transformation property
\lceq{ \label{eq:sec3_modular_property}
\Psi(M \langle Z \rangle ) = \chi (M) j(M, Z)^{c(0,0)/2}  \Psi(Z) \, ,\quad M \in \Gamma(L) \, .
}
$\chi(M)$ is called the multiplier system (see Definition~\ref{defn:sec3_multiplier_system}) and $\Psi(Z)$ is a modular form on generalized upper-half plane (also called Borcherds product) of weight $c(0,0)/2$ with the multiplier system (or character if the weight is integer) $\chi$ and modular group $\Gamma(L)$. This modular group contains some elements that do not preserve orientation. Since our symmetry group is $\SO(2,l;\R)$, the modular group we use is actually $\SG (L) := \Gamma(L) \cap \SO(L)$. We will be interested in the logarithm (the argument) of such modular forms. Hence we also need the information of its poles and zeros, where the argument at these points is not well defined. Remarkably, the positions of zeros and poles are totally determined by the principal part, consisting of all the terms with $n$ (in equation~(\ref{sec3:eq_nhmf})) negative
\lceq{\label{eq:sec3_pp_nhmf}
\sum_{\beta \in L'/L} \sum_{\substack{n \in \Z + q(\beta) \\ n < 0}} c(\beta,n) \mfe_\beta(n \tau) \, .
}
By Theorem 13.3 in \cite{Borcherds1998} (Theorem~\ref{Thrm:sec4_RQD}), zeros and poles of $\Psi(Z)$ lie in the divisor $(\Psi)$, which is the linear combinations of rational quadratic divisors $H(\beta,m)$ (Heegner divisors). The rational quadratic divisors $H(\beta,m)$ are  unions of orthogonal subspaces $H_\lambda$ with respect to the vector $\lambda \in \beta + L$, for $\beta \in L'/L$ and rational negative norm $m$,  
\lceq{
H_\lambda = \left\{[Z_L] \in \mathcal{K}^+| \, (Z_L, \lambda) = 0\right\} \, .
}
A rational quadratic divisor $H(\beta, m)$ is defined as
\lceq{
H(\beta, m) = \sum_{\substack{\lambda \in \beta + L \\ q(\lambda) = m}} H_\lambda  \, .
}
The zeros and poles of $\Psi(Z)$ are contained in the divisor $(\Psi)$ which is given by
\lceq{\label{eq:RQD}
    \left(\Psi\right) = \frac{1}{2} \sum_{\beta \in L'/L} \sum_{\substack{m \in \Z + q(\beta) \\ m<0}} c(\beta, m) H(\beta, m) \, .
}
These rational quadratic divisors are  closely related to the gauge symmetry enhancement  \cite{HARVEY1996315}. We will return to this in section~\ref{sec:general_strategy}.
 
\section{Composite\mathtitle{\U(1)}{U(1)}in minimal supergravity}\label{sec:U(1)_anomaly_in_8D_SUGRA_general}
We can now turn to computing the compensating $\U(1)$ transformation. As the first step a suitable parametrization of the coset space 
\lceq{
\mathcal{M} = \frac{\SO(2,l)}{\SO(2)\times \SO(l)} \cong \frac{\SO(2,l)}{\U(1)\times \SO(l)} 
}
is needed. The coset construction in \cite{1985_PLB_Salam} provides a good starting point. At the level of Lie algebra  the representative of the coset $\mathfrak{so}(2,l)/\left(\mathfrak{so}(2)\oplus \mathfrak{so}(l)\right)$ can be written as a matrix
\lceq{\label{eq:sec4_lie_algebra}
\mx{
0_{2\times 2} & H_{2 \times l} \\
(H^T)_{l \times 2} & 0_{l \times l} 
} \, , \quad H \in \text{Mat}(2 \times l, \R) \, .
}
Here the $2l$ real scalars $\phi^\alpha$ in $\mathcal{N}=1$ vector multiplets (see \eqref{eq:minsugra}) are packaged in $H$, and 
 $\text{Mat}(2\times l, \R)$ is the set of the real $2\times l$ matrices. An element $\Lambda \in \M$, $\Lambda^T \eta \Lambda =  \eta$,  can be represented as 
\lceq{\label{eq:sec4_vielbein}
\Lambda = \exp \mx{
0 & H\\
H^T &  0
} = \mx{
\sqrt{1+q q^T} & q \\
q^T & \sqrt{1 + q^T q} 
}\, ,
}
where $q \in M(2\times l, \R)$ is given by \lceq{
q = H\left(\frac{\sinh H^T H}{H^T H}\right)^{\frac{1}{2}} \, .
}
By direct matrix multiplication we can see that 
\lceq{
\sqrt{1 + q q^T} = \cosh (H H^T)^{1/2}, \quad \sqrt{1 + q^T q} = \cosh (H^T H)^{1/2}\, .
}
The matrices $\sqrt{1 + qq^T}$ and $\sqrt{1 + q^T q}$ satisfy the relation
\lceq{\label{eq:sec4_sqrtqTq}
\sqrt{1 + q^T q} = \id + q^T \left(\sqrt{1 + qq^T} - \id \right) (qq^T)^{-1} q \, ,
}
which can be checked by squaring the expression on the both sides. The negative power of the matrix in this expression should be considered in the sense of Taylor expansion since the matrix $H^T H$ might not be invertible. Based on this parametrization, we can further simplify the expression by introducing the so called  Calabi-Vesentini coordinates \cite{Calabi_1960_AM, kecker2009quaternion}. 

The matrix elements can be labeled by $\tensor{\Lambda}{_I^A}$, with $I$ being the row index and $A$ represents the column index. All the capital Latin indices take integer value from $0$ to $l+1$ ($I,A = 0, 1, \ldots, l+1$), and the metric-preserving property can be written in components as
\lceq{
\tensor{\Lambda}{_I}^A \tensor{\Lambda}{_J^B} \eta_{AB} = \eta_{IJ} \, .
}
The inverse matrix element of $\Lambda^{-1}$ is denoted as $\tensor{\Lambda}{^I_A}$ and satisfies
\lceq{
\tensor{\Lambda}{_I^A}\tensor{\Lambda}{^I_B} = \tensor{\delta}{^A_B} \, ,\quad\tensor{\Lambda}{^I_A} = \eta^{IJ}\eta_{AB} \tensor{\Lambda}{_J^B} \, .
}
We can now define 
\lceq{\label{eq:sec4_def_Phi}
\Phi^A = \frac{1}{\sqrt{2}} \left(\tensor{\Lambda}{_0^A} + i \tensor{\Lambda}{_1^A}\right),
}
which can be verified to satisfy
\lceq{\label{eq:Phi_cond}
\Bar{\Phi}^A \Phi^B \eta_{A B} = \frac{1}{2} \left(\tensor{\Lambda}{_0^A} - i \tensor{\Lambda}{_1^A} \right)\left(\tensor{\Lambda}{_0^B} + i \tensor{\Lambda}{_1^B}\right)\eta_{A B} = \frac{1}{2} (\eta_{00} + \eta_{11}) = 1\, , \\
\Phi^A \Phi^B \eta_{A B} = \frac{1}{2} \left(\tensor{\Lambda}{_0^A} + i \tensor{\Lambda}{_1^A} \right) \left(\tensor{\Lambda}{_0^B} + i \tensor{\Lambda}{_1^B}\right) \eta_{A B} = \frac{1}{2}(\eta_{00} - \eta_{11}) = 0  \, .
}
A natural Ansatz for $\Phi^A$ satisfying these constraints takes the form
\lceq{
\Phi^A = \frac{X^A}{\sqrt{\overline{X}^A X^B \eta_{AB}}} \, .
}
where $X^A$ are  components of a $l+2$ dimensional complex vector $\vec{X}$ such that $\vec{X}^T \eta \vec{X} = 0$.  
In terms of $X^A$ the matrix~(\ref{eq:sec4_vielbein}) can be written as 
\lceq{\label{eq:sec4_L(X)}
\Lambda = \frac{1}{\sqrt{2 \overline{X}^A X^B \eta_{AB}}} \mx{
X^0 + \bar{X}^0 & -i (X^0 - \bar{X}^0) & \ldots \\
X^1 + \bar{X}^1 & -i (X^1 - \bar{X}^1) & \ldots \\
\vdots & \vdots & * \\
X^{l+1} + \bar{X}^{l+1} & -i (X^{l+1} - \bar{X}^{l+1}) 
} \, .
}
Notice that $-i(X^0 - \bar{X}^0) = X^1 + \bar{X}^1$, $\Lambda$ is a symmetric real matrix.

One way to parametrize $X^A$ in terms of of $l$ independent complex scalars 
is
\lceq{\label{eq:sec4_CV_coordinate}
X^A = \left(\frac{1+y^2}{2}, \frac{ i}{2}(1-y^2), y_i\right), \quad i = 1, \ldots, l \, ,
}
where $y_i$ is a complex scalar and $y^2 := y_i y_i$. Here, and in the rest of the discussion, a  summation over all repeated indices is implied. In addition, $y_i$ should satisfy \cite{Calabi_1960_AM}
\lceq{\label{eq:sec4_y_constraint}
\overline{X}^A X^B \eta_{AB} > 0 \quad \Rightarrow \quad 1 - 2 \by_i \by_i + y^2 \by^2 > 0, \quad \by_i y_i < 1\, ,
}
which is the bounded choice of the region for $y_i$, known as Calabi-Vesentini coordinates.  In terms of $y^i$ 
\lceq{\label{eq:sec4_L(y)}
\Lambda = \frac{1}{\sqrt{\deno}} \left[
\begin{matrix}
1+ \frac{1}{2}\left(y^2 + \by^2 \right) & -\frac{i}{2} (y^2 - \by^2) & \ldots \\
-\frac{i}{2} (y^2 - \by^2)  & 1 - \frac{1}{2} (y^2 + \by^2) & \ldots \\
y_1 + \by_1 & -i(y_1 - \by_1)  \\
\vdots &\vdots & * \\
y_l + \by_l & -i(y_l - \by_l) &
\end{matrix}
\right].
}
Under the parametrization~(\ref{eq:sec4_L(X)}), it is easy to see that $\vec{X}$ is equivalent with $t  \vec{X}$ if $t \in \R$. Also, recall that $\Lambda$ is a coset representative, i.e. $\Lambda \sim \Lambda U$ where $U$ is a $\SO(2)\times \SO(l)$ transformation parametrized by a real  $\theta$:
\longeq{
\Lambda \sim \Lambda U &= \frac{\sqrt{2}}{\sqrt{\vec{X}^\dg \eta \vec{X}}} \mx{
\Re \left(\vec{X}\right) & \Im \left(\vec{X}\right) & *
} \mx{
\cos \theta & -\sin \theta &  \\
\sin \theta & \cos \theta &  \\
& & U_{\SO(l)} 
} \\
&= \frac{\sqrt{2}}{\sqrt{\vec{X}^\dg \eta \vec{X}}} \mx{
\Re \left(\vec{X}e^{-i \theta}\right) & \Im \left(\vec{X}e^{-i \theta}\right) & *
} \, ,
}
which means that $\vec{X} \sim \vec{X}e^{-i \theta}$. Combined with the scaling transformation, this leads to a conclusion that $\vec{X}$ lives in the projective space and $\vec{X} \sim \alpha \vec{X}$ for an arbitrary non-zero complex number $\alpha$. 

\subsection{\mathtitle{\U(1)}{U(1)}connection, gauge transformations and gauge fixing}\label{subsec:U(1)connection}
We are now ready to construct explicitly the composite connection associated with the local $\U(1)$ gauge symmetry. It can be expressed in terms of the Maurer-Cartan form \cite{1985_PLB_Salam} as
\lceq{
Q = \tensor{\left(\Lambda^{-1} d \Lambda\right)}{_0^1}, \quad Q_\mu = \tensor{\left(\Lambda^{-1} \pl_\mu \Lambda\right)}{_0^1}\, ,
}
where $d$ is the exterior derivative defined on the spacetime manifold and $\pl_\mu$ is the partial derivative with respect to the $\mu$ spacetime coordinates. Using the expression of  $\Lambda$  of $q$ (\ref{eq:sec4_vielbein}), we  have 
\lceq{
(\Lambda^{-1} \pl_\mu \Lambda)_{2 \times 2} = \sqrt{1+q q^T} \pl_\mu \sqrt{1+q q^T} - q \pl_\mu q^T \, ,
}
where the subscript indicates the $2\times 2$ upper left corner of the matrix $\Lambda^{-1} \pl_\mu \Lambda$. Since $y_i$ are unconstrained variables, expressing $Q$ in terms of these avoids ambiguities and we have:
\longeq{\label{eq:sec4_Qmu}
Q_\mu  =  2i \frac{\by_i - \by^2 y_i}{\deno} \pl_\mu y_i - \frac{i}{2} \pl_\mu \ln \left( \deno\right) \, .
}
Notice that 
\lceq{
 \frac{1}{2} \left(\deno\right) =  \bar{X}^A X^B \eta_{AB} = \vec{X}^\dg \eta \vec{X} \, ,
}
the denominator is naturally invariant under the transformation $\vec{X} \rightarrow \vec{X} e^{-i \Sigma}$. Besides, 
\longeq{
\vec{X}^\dg \eta \pl_\mu \vec{X} &= \left(\frac{1 + \by^2}{2}, -\frac{i}{2} (1 - \by^2), \vec{y}^\dg   \right) \mx{
\id & 0 \\
0 & -\id
} \mx{
y_i \pl_\mu y_i \\
-i y_i \pl_\mu y_i \\
\pl_\mu \vec{y}
} \\
&= \by^2 y_i \pl_\mu y_i - \vec{y}^\dg \pl_\mu \vec{y} 
}
allows to  write $Q_\mu$ compactly in terms of $\vec{X}$,
\lceq{
Q_\mu  = -i \frac{\vec{X}^\dg \eta \pl_\mu \vec{X}}{\vec{X}^\dg \eta \vec{X}} - \frac{i}{2} \pl_\mu  \ln \left( 2 \vec{X}^\dg \eta \vec{X}\right) \, .
}
Under the $\U(1)$ gauge transformation $
\vec{X} \rightarrow \vec{X}' = \vec{X}e^{-i \Sigma}$ we have 
\lceq{
\delta Q_\mu = Q'_\mu - Q_\mu  = -i \frac{\vec{X'}^\dg \eta \pl_\mu \vec{X'}}{\vec{X'}^\dg \eta \vec{X'}} + i \frac{\vec{X}^\dg \eta \pl_\mu \vec{X}}{\vec{X}^\dg \eta \vec{X}} = -  \pl_\mu \Sigma \, ,
}
as expected ($Q'_\mu$ here denotes the  $U(1)$ transformed connection).

Notice that in the expression for the  coset element $\Lambda$~(\ref{eq:sec4_vielbein}) we have chosen the gauge $\phi = 0$, where $\phi$ represents the variable parametrizing local $\U(1)$ gauge symmetry. In order to maintain the gauge ($\phi = 0$), a left action on $\Lambda\in \mathcal{M}$ by an $\SO(2,l)$ transformation should be compensated by a right action of a $\SO(2)\times \SO(l)$ transformation, i.e.
\lceq{\label{eq:sec4_action_coset}
\Lambda \rightarrow \Lambda' = R \Lambda U^{-1}, \, \, U \in \SO(2) \times \SO(l) \, ,
}
which leads to 
\longeq{
\frac{1}{\sqrt{\vec{X}^\dg \eta \vec{X}}}R \mx{
\Re \left(\vec{X}\right) & \Im \left(\vec{X}\right) 
} &= \frac{1}{\sqrt{\vec{Y}^\dg \eta \vec{Y}}} \mx{
\Re \left(\vec{Y}\right) & \Im \left(\vec{Y}\right) 
}\mx{
\cos \Sigma & -\sin \Sigma &  \\
\sin \Sigma & \cos \Sigma 
} \\
&= \frac{1}{\sqrt{\vec{Y}^\dg \eta \vec{Y}}} \mx{
\Re \left(\vec{Y}e^{-i \Sigma}\right) & \Im \left(\vec{Y} e^{-i \Sigma}\right)
} \, ,
}
where the complex vector $\vec{Y}$ parametrizes the new coset representative $\Lambda'$. The equation is obviously equivalent to 
\lceq{\label{eq:sec4_U(1)_gauge_fixing}
\frac{1}{\sqrt{\vec{X}^\dg \eta \vec{X}}} R \vec{X} = \frac{1}{\sqrt{\vec{Y}^\dg \eta \vec{Y}}} \vec{Y} e^{-i \Sigma} \, .
}
In the next subsection we will proceed to formally solving this equation and obtaining an analytic expression of $\Sigma = \Sigma(R,\vec{X})$. Before doing so we should recall that in the familiar case of $\SL(2,\R)/\U(1)$, the compensating $\U(1)$ transformation (the phase factor) is given by \eqref{eq:sec1_U(1)trans} in terms of the modular variable $\tau$ which lives in the complex upper-half plane $\mathbb{H}$.  As we shall see, the relation between the $\U(1)$ anomaly and modular variables is universal. 

\subsection{Compensating U(1) transformation}
We start by recalling that the vector $\vec{X}$, which lives in the projective space, satisfies the constraints 
\lceq{
\vec{X}^T \eta \vec{X} =  0 , \quad \vec{X}^\dg \eta \vec{X} > 0 \, .
}
This  matches the condition~(\ref{eq:sec4_mathcalK}) on the generalized upper-half plane for the group $\mO^+(2,l; \R)$ (see section~\ref{subsec:sec4_action_O_on_generalized_plane}). The Calabi-Vesentini coordinates~(\ref{eq:sec4_CV_coordinate}) are not very convenient for solving the equation~(\ref{eq:sec4_U(1)_gauge_fixing}) and determining $\Sigma(R,\vec{X})$. Instead, we should rotate to the reference frame with basis already discussed in section~\ref{sec:Mathematical_preliminaries}, and notably use matrices $M \in \mO^+(2,l;\R)_{S}$. The subscript here emphasizes that the orthogonal group is defined with respect to a metric $S$. This definition applies throughout our discussion, and we shall often omit the subscript.
 
 Since $\hat{S}$, defining the metric $S$, introduced in \eqref{eq:sec4_qudratic_form} is a symmetric positive-definite real matrix, there must exist a orthogonal matrix $\hat{P}$ such that $ \hat{P} \hat{S} \hat{P}^T = \hat{V}$, where $\hat{V}$ is the diagonal matrix with positive diagonal elements. One can define the square root of the inverse $\sqrt{\hat{V}^{-1}}$ such that 
\lceq{
 \id = \hat{Q} \hat{S}  \hat{Q}^T  \, , \quad \hat{Q} = \sqrt{\hat{V}^{-1}} \hat{P} \, .
}
There exists a orthogonal matrix $U$ such that $U S U^T = \eta$, explicitly we have  
\lceq{
U = \mx{
\frac{1}{\sqrt{2}} J &  & \frac{1}{\sqrt{2}} \id_2  \\
 & \hat{Q} &  \\
-\frac{1}{\sqrt{2}} J & & \frac{1}{\sqrt{2}} \id_2 
}, \quad J = \mx{
0 & 1\\
1 & 0
} , \quad  U U^T  =  \mx{
\id_2 & & \\
& \hat{V} & \\
& & \id_2
} \, .
}
Inserting $U$ into the equation~(\ref{eq:sec4_U(1)_gauge_fixing})  we have 
\lceq{\label{eq:sec4_U(1)_gauge_fixing_mid}
\frac{1}{\sqrt{\vec{Z}^\dg S \vec{Z}}} M \vec{Z} = \frac{1}{\sqrt{\vec{W}^\dg S \vec{W}}} \vec{W} e^{-i \Sigma}  , \quad \vec{Z} = U^T \vec{X}, \quad \vec{W} = U^T \vec{Y} \quad M = U^T R (U^T)^{-1} \, .
}
It is not difficult to verity that $M \in \mO^+(2,l;\R)_{S}$. To further demonstrate that such a choice of  coordinates would be realized as the generalized upper-half plane, we explicitly expand the equation and derive the constraints satisfied by $\vec{Z}$ \cite{gilmore2012lie}. After the rotation we have 
\lceq{\label{eq:sec4_beta}
\vec{Z} = U^T \vec{X} = \mx{
\frac{i}{2 \sqrt{2}} (1 - y^2) - \frac{1}{\sqrt{2}} y_{s+2} \\
\hdashline 
\frac{1}{2\sqrt{2}} (1+y^2) - \frac{1}{\sqrt{2}} y_{s+1} \\
\hat{Q}^T \vec{y}_s \\
\frac{1}{2\sqrt{2}} (1+y^2) + \frac{1}{\sqrt{2}} y_{s+1} \\
\hdashline
\frac{i}{2 \sqrt{2}} (1 - y^2) + \frac{1}{\sqrt{2}} y_{s+2}
}  = \mx{
\beta_0 \\
\hdashline 
\beta_1 \\
\vdots  \\
\beta_{s+2} \\
\hdashline
\beta_{s+3}
} \, ,
}
satisfying the constraint
\longeq{\label{eq:sec5_constraintZ}
\begin{cases}
\vec{Z}^T S \vec{Z} = 0 &\Rightarrow \quad 2 \beta_0 \beta_1 + 2 \beta_{s+2} \beta_{s+3} - \vec{\beta}_s^T \hat{S} \vec{\beta}_s = 0 , \\
\vec{Z}^\dg S_1 \vec{Z} > 0 &\Rightarrow \quad  \overbar{\beta_0} \beta_1 + \overbar{\beta_1} \beta_0 - \vec{\beta}_s^\dg \hat{S} \vec{\beta}_s + \overbar{\beta_{s+2}} \beta_{s+3} + \beta_{s+2} \overbar{\beta_{s+3}} > 0 \, ,
\end{cases} 
}
where $\vec{\beta}_s$ is the vector with components $(\beta_2, \beta_3, \ldots, \beta_{s+1})$.  
First, let us verify that $\beta_{s+3}\neq 0$. Indeed, assuming $\beta_{s+3} = 0$, would yield
\lceq{
\begin{cases}
2 \beta_0 \beta_1 -  \vec{\beta}_s^T \hat{S} \vec{\beta}_s = 0 , \\
\overbar{\beta_0} \beta_1 + \overbar{\beta_1} \beta_0  - \vec{\beta}_s^\dg \hat{S} \vec{\beta}_s > 0 .
\end{cases}
}
However, by absolute inequality, we have \lceq{
\left|2 \beta_0 \beta_1\right| = |\vec{\beta}_s^T \hat{S} \vec{\beta}_s | \le  \vec{\beta}_s^\dg \hat{S} \vec{\beta}_s < \overbar{\beta_0} \beta_1 + \overbar{\beta_1} \beta_0 \, .
}
Since both sides of the equation have positive signs, we can square  without changing  the direction of the inequality:
\lceq{
4 \overbar{\beta_{0}} \beta_0 \overbar{\beta_1} \beta_1 < \left(\overbar{\beta_0} \beta_1 \right)^2 + \left(\overbar{\beta_1} \beta_0\right)^2 + 2 \overbar{\beta_0} \beta_0 \overbar{\beta_1} \beta_1 \quad \Longleftrightarrow \quad \left(\Im \overbar{\beta_0} \beta_1\right)^2 < 0 \, ,
}
leading to contradiction. Thus $\beta_{s+3} \neq 0$ and we can safely normalize the vector $\vec{Z}$ by dividing the final component, 
\lceq{\label{eq:sec4_z}
\vec{Z} = \alpha(z) \mx{
-q_0(Z) \\
Z \\
1
} = \alpha(Z) Z_L, \, \alpha(Z) = \beta_{s+3}, \, Z\in \C^{s+2}, \, Z_j = \frac{\beta_{j}}{\beta_{s+3}}, \, j = 1, \ldots , s+2 .
}
Here $q_0(Z) = Z^T S_0 Z/2$ as defined in section~\ref{subsec:sec4_action_O_on_generalized_plane}. With the definition of the quadratic form $q(Z_L) = \frac{1}{2} Z_L^T S Z_L$ and $(A,B) = q (A+B) - q (A) -q(B)$, we can rewrite the constraints~(\ref{eq:sec5_constraintZ}) of $Z_L$ as
\lceq{
\left(Z_L, Z_L\right) = 0 \, ,\quad \left(Z_L, \overbar{Z_L}\right) > 0 \, , 
}
so we conclude that $Z_L \in \mathcal{K}$ defined by equation~(\ref{eq:sec4_mathcalK}).  Without loss of generality we assume that $Z_L \in \mathcal{K}^+$, and can check that  the only constraint on th erange of $Z$ is given by  $q_{0} \left(\Im(Z)\right) > 0 $. If we assume $\Im (Z)$ lives in the future light cone of the Minkowski space, $Z$ indeed lives in the generalized upper-half plane $\mH_l$. With these setting we can rewrite the equation~(\ref{eq:sec4_U(1)_gauge_fixing_mid}) as 
\lceq{
\frac{e^{i \hat{\phi}(Z)}}{\sqrt{Z_L^\dg S Z_L}} M Z_L = \frac{e^{i \hat{\phi}(W)}}{\sqrt{W_L^\dg S W_L}} W_L e^{-i \Sigma}, 
}
where $Z_L = \left(-q_0(Z), Z ,1\right)^T$, $W_L = \left(-q_0(W), W, 1\right)^T$, and 
\lceq{
e^{i \hat{\phi}(Z) }= \frac{\alpha(Z)}{|\alpha(Z)|}\, , \quad e^{i \hat{\phi}(W) = }\frac{\alpha(W)}{|\alpha(W)|} \, .
}
Recalling the discussion of the action of the orthogonal group on generalized upper-half plane (equation~(\ref{eq:sec4_action_M2}) and~(\ref{eq:sec4_action_M3})), we conclude that $W = M\langle z \rangle$ and 
\lceq{
e^{-i \Sigma(M,Z)} = e^{i \hat{\phi}(Z) - i \hat{\phi}(W)} \frac{\sqrt{W_L^\dg S W_L}}{\sqrt{Z_L^\dg S Z_L}} \left(-\gamma q_0(Z) + d^T Z + \delta\right) \, .
}
Recall that 
\lceq{
M Z_L = \mx{
-\alpha q_0(Z) + a^T Z + \beta \\
- b q_0(Z) + P Z + c \\
-\gamma q_0(Z) + d^T Z + \delta 
} = \left(-\gamma q_0(Z) + d^T Z + \delta \right) W_L \, ,
}
with the property that the real orthogonal transformation doesn't change the norm, i.e. $\sqrt{Z_L^\dg S Z_L} = \sqrt{\left(M Z_L\right)^\dg S \left(M Z_L\right)} $, we conclude that
\lceq{
 e^{-i \Sigma(M,Z)} = 
 e^{i \hat{\phi}(Z) - i \hat{\phi}(W)} \frac{-\gamma q_0(Z) + d^T Z + \delta}{\left|-\gamma q_0(Z) + d^T Z + \delta\right|}  = e^{i \hat{\phi}(Z) - i \hat{\phi}(W)} \frac{j(M,Z)}{|j(M,Z)|} \, .
 }
 By choosing the specific gauge, the compensating $\U(1)$ transformation is given by
 \lceq{\label{eq:sec5_compensatingU(1)}
 e^{-i \Sigma(M,Z)} = \frac{j(M,Z)}{|j(M,Z)|} \, .
}
This is the direct generalization of the compensating $\U(1)$ transformation for $\SL(2,\R)/\U(1)$ given in \eqref{eq:sec1_U(1)trans} to the generalized upper-half plane $\mathbb{H}_l$.  
\section{Constructing the counterterm}
\label{sec:general_strategy}
As already discussed in the beginning of section~\ref{sec:Mathematical_preliminaries}, eight-dimensional $\mathcal{N}=1$ theories suffer from a composite $\U(1)$ anomaly. The anomalous phase raised in the local $\U(1)$ gauge transformation $(\phi \rightarrow \phi +\Sigma)$ is
\lceq{\label{eq:anno}
\Delta_G = - \int \Sigma X_8(R ,\mathcal{F}) \, .
}
A direct way to cancel the anomalous phase is to add the local counterterm
\lceq{
\mathcal{S}_\phi = \int \phi X_8(R,\mathcal{F}) \, ,
}
where $\phi$ parametrizes the local $\U(1)$ gauge symmetry. When we apply the $\U(1)$ gauge transformation $\phi \rightarrow \phi + \Sigma$, $\delta \mathcal{S}_\phi$ can cancel the anomalous phase above. But the drawback is that the local counterterm is not invariant under $\SO(2, l ;\R)$ symmetry transformations, as shown in the equation~(\ref{eq:sec5_compensatingU(1)})
\lceq{
\delta_M \phi = - \arg \left(j(M,Z)\right) \, .
}
Here $\delta_M$ indicates an $\SO(2,l;\R)$ gauge transformation with respect to the element $M$. Since the compensating $\U(1)$ transformation is the argument of the automorphy factor, it is natural to construct the counterterm by using modular forms on generalized upper-half plane and it is of the form
\lceq{\label{eq:cooo}
\mathcal{S} =  \frac{1}{r} \int \arg (\Psi(Z)) X_8 (R,\mathcal{F}) \, ,
}
where $\Psi(Z)$ satisfies the modular property~(\ref{eq:modular_form}). We have already seen that the continuous symmetry group should be discretized since no suitable functions that can maintain the continuous symmetry. As mentioned in section~\ref{sub:orthogonal_modular_forms} the analogue of  $\SL(2;\Z)$  in $\mathcal{N}=2$ case is  the discrete modular group $\SG(L)$ with respect to the lattice $L$ of signature $(2,l)$. Such a discrete lattice $L$ will be the root lattice of the gauge group $G$ (or contain the sublattices which are the root lattices of the gauge group $G$). Hence the anomaly cancellation may lead to nontrivial restrictions on the lattices $L$ (on the gauge groups $G$). 

As discussed in section~\ref{sec:Mathematical_preliminaries},  the Borcherds products provide necessary tools for constructing $\Psi(Z)$ with requisite properties to cancel the anomaly \eqref{eq:anno}. As long as a nearly holomorphic modular form of weight $1 - l /2$ with respect to the lattice $L$ can be found, one can obtain the modular form $\Psi(Z)$ on generalized upper-half plane of weight $r = c(0,0)/2$. However the counterterms needs to satisfy some natural physical conditions leading to constraints that will be outlined bellow.\footnote{Here we recall another time that throughout this discussion we have taken the lattice $L$ to be even.}
\begin{itemize}[leftmargin=*]
  \item \textbf{The character of the lattice $L$ (the modular group $\SG(L)$)}
\end{itemize}
Since the modular form $\Psi(Z)$ satisfies the modular property~(\ref{eq:sec3_modular_property}), where the weight $r = c(0,0)/2$, the counterterm is transformed under the $\SG(L)$ transformation as
\lceq{ 
\delta_M \mathcal{S} = - \delta_M \mathcal{S}_{\phi}   +  \arg \chi(M) \int   X_8(R,\mathcal{F})\, .
}
In order to completely cancel the anomaly without imposing extra conditions on the background manifold, such as integrality of $\int X_8 (R,\mathcal{F})$, some extra conditions need to be imposed on the lattice $L$ or the modular group $\SG(L)$. There are two possible ways to resolve this problem:
\begin{enumerate}
    \item $\chi(M) \equiv 1$ for arbitrary $M \in \SG(L)$ is required. 
    \item The modular group $\SG(L)$ breaks to its subgroup $\widetilde{\SG} (L)$ such that for arbitrary $M \in \widetilde{\SG} (L)$ we have $\chi(M) \equiv 1$. 
\end{enumerate}
\noindent
To the best of our knowledge,  the necessary and sufficient condition for the character to be trivial is not known.\footnote{In general $\chi(M)$ is called the multiplier system and  is different from character if the weight of the Borcherds product is not integral. Through suitable normalization we can always obtain the Borcherds product of integral weights so we will not consider the cases of rational weights.} 
A sufficient condition is known (Theorem \ref{Thrm:sec4_trivial_character}). Moreover, it cannot be weakened too much (see the counter example (Example 1.4) in \cite{GRITSENKO2009JA}). More details are in the appendix~\ref{app:supplement_modular_form}. Notably any lattice that contains $A_2$ sublattice has $\chi(M) \equiv 1$.

\begin{itemize}[leftmargin=*]
  \item \textbf{Rational quadratic divisor (RQD)}
\end{itemize}

The counterterm is obviously ill-defined at the zeroes or poles of $\Psi(Z)$.  To circumvent this issue, one could have required  the Borcherds product $\Psi(Z)$ to be well-defined and have no zeroes on the entire generalized upper-half plane, which is equivalent to requiring $c(\beta, m ) \equiv 0$ if $m < 0$ for all $m \in \Z + q(\beta)$ and $\beta \in L'/L$. This would in turn mean that the principal part of the nearly holomorphic modular form $f(\tau)$ is zero, so $f(\tau)$ is actually a holomorphic modular form of $\SL(2,\Z)$. However, no nonzero holomorphic modular forms of non-positive weight ($1 - l /2 \le 0$ for $l \ge 2$) exist and thus the counterterm will always have ill-defined points in moduli space.

In fact, a nearly holomorphic modular form $f(\tau)$ exists if and only if  the coefficients in the principal part~\eqref{eq:sec3_pp_nhmf}  satisfy
\lceq{\label{eq:RefL}
\sum_{\beta \in L'/L}  \sum_{\substack{m \in \Z +q (\beta) \\ m<0}} c(\beta, m) a_{\beta, -m} \equiv 0 , \, 
}
where $a_{\beta, -m}$ is the functional that maps the cusp form $g \in S_{\kappa, L}$ into its $(\beta,-m)$ Fourier coefficient and $S_{\kappa,L}$ is the space of the cusp forms of weight $\kappa = 1 + l/2$ for the dual Weil representation (more details can be found in the discussion of Theorem~\ref{Thrm:appB_serre_dual}). 

The simplest solution to this condition is when the cusp form space $S_{\kappa, L}$ is trivial (there exists no nonzero cusp form of weight $1+l/2$ of dual Weil representation). Lattices with such property exist and are called simple lattices. As shown in \cite{Dittmann2015}, there are only 15 simple even lattices of signature $(2,l)$, $l \ge 4$, of square free level\footnote{The level of the lattice $L$ is a positive integer $p$ such that $p = \min\{n \in \N| \, n q(\gamma) \in \Z \, \text{for all}\, \gamma \in L'\}$.} up to isomorphisms (see the Theorem 2 in \cite{Dittmann2015}).  For signature $(2,18)$, only the even unimodular lattice $\Pi_{2,18} \cong \Pi_{1,1} \oplus \Pi_{1,1} \oplus E_8(-1)\oplus E_8(-1)$ is simple, and for $(2,10)$ only the even unimodular  $\Pi_{2,10} \cong  \Pi_{1,1} \oplus \Pi_{1,1} \oplus E_8(-1)$ and $\Pi_{1,1} \oplus \Pi_{1,1}(2) \oplus E_8(-1)$ are simple. 

From Theorem~\ref{Thrm:sec4_RQD} we know that all the zeroes and poles of Borcherds products lie in the rational quadratic divisors (Definition~\ref{defn:sec3_RQD}). These ill-defined points in moduli space correspond to symmetry enhancement (as pointed out in the context of 4D $\mathcal{N} = 2$ theories in \cite{HARVEY1996315}). RQDs form the set of the orthogonal subspaces determined by the negative-norm vectors $\ell \in L'$,  such that the reflections orthogonal to them are symmetries of the lattice. This reflection symmetry leads to the appearance of extra massless states needed for the symmetry enhancement.  Consequently, we require the lattice $L$ defining the eight-dimensional theory to be reflective.

\begin{itemize}[leftmargin=*]
  \item \textbf{Reflective lattices} 
\end{itemize}
There is a finite number of reflective lattices, and their rank is bounded by $l=26$. There is a complete classification of reflective lattices of prime level. All these lattices are of even rank and hence should be considered. A complete classification for any level  is available for a particular subclass, the $2$-reflective lattices that have norm $-2$ roots.\footnote{A bibliographical note: In \cite{Scheithauer2006} all strongly reflective modular forms of singular weight on lattices of prime level were classified. A proof that there are only finitely many even lattices with $l \ge 7$ which admit $2$-reflective modular forms and the highest rank such lattice is the the even unimodular lattice $\Pi_{2,26}$ is given by \cite{ma2017}. These were subsequently classified in \cite{Wang2023IMRN, wang2023JEMS, wang2019IMRN}. In \cite{wang2022}, all possible reflective lattices of prime level were classified. In our discussion of reflective modular forms (see subsection  \ref{subsec:reflective_mf_lattice}), we will adopt the conventions of \cite{wang2022}. } Here we find lattices of odd rank, which should be discarded due to the global anomalies.

Recall that in ten dimensions anomaly cancellation allows for not only rank 16 theories (with a unimodular lattice $E_8 \oplus E_8$), but also of theories with gauge group $E_8 \times \U(1)^{248}$ and $\U(1)^{496}$. Simple reduction of these theories would produce 8D theories with $l=258$ and $l=498$ respectively. The fact that the condition of reflectivity bounds the rank of the lattice to be equal or less than 24 tells us that for these no suitable 8D counterpart can be found (even if they admit 10D Green-Schwarz term). So it seems  these theories can be ruled out purely based on anomaly cancellation, and without swampland considerations.

\begin{itemize}[leftmargin=*]
  \item \textbf{Counterterms as obstructions to ten-dimensional lifts} 
\end{itemize}
Given the form of a reflective lattice \eqref{eq:reflL} it is natural to ask about possible decompactifications to ten dimensions. If such decompactification is possible , i.e. a good ``large volume limit'' exists, the 8d theory can be considered consistent only if a lifting to the ten-dimensional $E_8 \oplus E_8$ lattice exists.\footnote{For $L \cong \Pi_{1,1} \oplus \Pi_{1,1} \oplus \sum_i \hat{L}_i$, there always exists a straightforward lift in ten dimensions with $L(10D) = \sum_i \hat{L}_i$. Other than for the Narain lattice, all these lifts can be discarded. The CHL lattice $\Pi_{1,1} \oplus \Pi_{1,1} \oplus D_8(-1) \cong \Pi_{1,1} \oplus \Pi_{1,1}(2) \oplus E_8(-1)$ also allows for a lift to $E_8 \oplus E_8$.  Very loosely, all lattices that would allow to have central charge $c_L = 18$ can be potentially liftable to ten dimensions}

We have not done an exhaustive check on which reflective lattices can or cannot be lifted to an $E_8 \oplus E_8$ lattice in 10D. Any lattice with $l >18$ clearly does not have such lifting. The rank 8 self-dual lattice  $\Pi_{1,1} \oplus \Pi_{1,1} \oplus E_8(-1)$ also does not have such lifting. 
For such lattices the anomaly cancellation can be validated only if they are ``intrinsically eight-dimensional'', i.e. if their counterterm obstructs the decompactification to 10D\footnote{Similarly, the function $F^{\text{het}}_1$ that appears in the one-loop gravitational couplings in the $\mathcal{N}=2$ heterotic compactification with two vector multiplets also does not allow such identification \cite{Kaplunovsky:1995tm}.}.

\subsection{Reflective modular forms and reflective lattices}
\label{subsec:reflective_mf_lattice}
Let $L$ be an even lattice of signature $(2,l)$ and its dual is $L'$. The level of $L$ is the smallest positive integer $N$ such that $N (x,x) \in 2 \Z$ for all $x \in L'$. The discriminant of $L$ denoted $L'/L$ can be decomposed by Jordan components and we denote this decomposition by $D_L$. The genus of $L$ is the set of lattices which have the same signature and the same discriminant form (up to isomorphism) as $L$. A holomorphic modular form for the modular group $\Gamma(L)$ is called reflective if its zeroes are contained in the union of rational quadratic divisors $\ell^\perp$ associated to roots of $L$, namely the reflection
\lceq{
\sigma_\ell: \alpha \longmapsto \alpha - \frac{2 (\alpha, \ell)}{(\ell,\ell)} \ell \, ,\quad \alpha \in L \, 
}
belongs to $\mO^+(L)$. A lattice is called reflective if it has a reflective modular form. A modular form is called symmetric if it is modular for $\mO^+(L)$ and it is known that $L$ is reflective if and only if $L$ has symmetric reflective modular form. Recall the definition of the modular group $\Gamma(L)$
\lceq{
\Gamma(L) = \mO^+(L \otimes \R) \cap \text{Ker} \left(\mO(L) \rightarrow \mO(L'/L)\right) \, , 
}
It is reasonable to require the local counterterms are constructed by symmetric modular form since we want to maintain the discrete symmetry maximally. 

We consider the lattices of the same genus $\Pi_{2,l}(p^{\epsilon_p l_p})$, where $l \ge 3$, $p$ is a prime number, $\epsilon_p = - \, \text{or} \, +$, $1 \le l_p \le l/2+1$ and $\epsilon_p$ is completely determined by $l,p$ and $l_p$. If two lattices of signature $(2,l)$ and prime level $p$ have the same determinant then they are isomorphic. We refer the readers to \cite{Scheithauer2006} for more details. Let $L$ be such a lattice.  By \cite{VVNikulin1980}, $L$ can be represented as\footnote{For some lattices, such as CHL, both representations are possible. Of course the last factor in two different ways of representing the lattice will also be different.} 
\lceq{\label{eq:reflL}
\Pi_{1,1} \oplus \Pi_{1,1}(p) \oplus \hat{L}(-1) \quad \text{or} \quad  \Pi_{1,1} \oplus \Pi_{1,1} \oplus \hat{L}(-1) \, ,
}where $\Pi_{1,1}$ is a hyperbolic plane as we defined above and $\hat{L}$ is a positive definite lattice. A primitive vector $v \in L$ is reflective if and only if $(v,v) = -2$ or $(v, v) = -2 p$ and $v / p \in L'$. By \cite{Scheithauer2015} and Eichler criterion (see e.g. \cite{gritsenko2010reflective}) all the vectors of norm $-2$ in $L$ are in the same $\mO^+(L)$-orbit, and all reflective vectors of norm $-2p$ in $L$ are also in the same $\mO^+(L)$-orbit. Therefore, for a symmetry reflective modular form, all $2$-reflective divisors (the rational quadratic divisors defined by the vector $v$ of norm $-2$) have the same multiplicity, which is denoted by $c_1$. All $2p$-reflective divisors (the rational quadratic divisors defined by the vector $v$ of norm $-2p$ and $v/p\in L'$) have the same multiplicity denoted by $c_p$. A symmetric reflective modular form is called $2$-reflective (resp. $2p$-reflective) if $c_p = 0$ (resp. $c_1$ = 0). A lattice $L$ is called $2$-reflective (resp. $2p$-reflective) if it has a $2$-reflective (resp. $2p$-reflective) modular form. 

The positions of the zeroes and poles of the modular form $\Psi(Z)$, where the counterterm \eqref{eq:cooo} is ill-defined, should be interpreted as the symmetry enhancement points. These points corresponds to  the rational quadratic divisors, which are defined as the orthogonal subspace with respect to some negative norm vectors (roots of the lattices). The symmetry is enhanced due to the reflective symmetry of the lattice. Requiring that $\Psi(Z)$ is symmetric reflective modular form, and thus the corresponding lattice $L$ should be reflective, ensures that the theory is well-defined and anomaly-free throughout the moduli space. This is a strong constraint for the lattice. As shown in  \cite{wang2022}, only  $55$ possible types of reflective lattices of genus $\Pi_{2,l}(p^{\epsilon_p l_p})$ with $1 \le l_p \le 1+ l/2$ exist for prime level $p >1$. And only three even unimodular lattices ($p=1$) $\Pi_{2,10}$, $\Pi_{2,18}$ and $\Pi_{2,26}$ are reflective. Among these lattices, only those with trivial character (a big majority) can provide suitable counterterms \eqref{eq:cooo} and hence lead to theories that are  anomaly-free. Further restrictions may be imposed by the consistency of the large volume limits. Since only the Narain and the CHL lattice pass the tests of full quantum consistency, all other lattices which lead to anomaly-free theories constitute the finite swampland of the eight-dimensional minimal supergravity.

\subsection{Examples of counterterms}
Before turning to specific examples of counterterms, we point out that if we choose the divisor of $\Psi(Z)$ to be a linear combination of some rational quadratic divisors, $\Psi(Z)$ must be the same function up to normalization as we construct from Borcherds product. More precisely (see Theorem 1.2 in \cite{Bruinier2014}), assuming that $L \cong \Pi_{1,1} \oplus \Pi_{1,1}(N) \oplus \hat{L}(-1)$ for some positive integer $N$ and $l \ge 3$,  every meromorphic modular form $F(Z)$ with respect to $\Gamma(L)$ whose divisor is a linear combination of special divisors $H(\beta, m)$ is (up to a non-zero constant factor) the Borcherds product $\Psi(Z)$ of some $f \in M^!_{1 - l/2}$. 

We can now discuss examples, which include two fully consistent 8D $\mathcal{N}=1$ supergravities with $l=18$ and $l=10$.

\begin{itemize}[leftmargin=*]
  \item \textbf{Signature $(2,18)$} 
\end{itemize}
As already mentioned reflectivity imposes an upper bound $l=26$ on the rank of the gauge group. Moreover, for 
$l > 18$ there are very few reflective lattices with $l$ even: the self-dual lattice $\Pi_{2,26}$ and two lattices at level $2$ and $3$, $\Pi_{2,22}(2)$ and $\Pi_{2,20}(3)$ respectively. If these allow a decompactification limit, they can be ruled out.

The $l = 18$ case, in addition to the self-dual (Narain) lattice, includes five different level $2$-reflective lattices.
These five will necessarily have enhancement points corresponding to norm $-4$ root vectors.  Their modular forms cannot be decomposed into products of $2$-reflective and $4$-reflective forms, and they have no string theory realization. 

For the theory obtained via compactification of $10$D heterotic string  on a two-torus  \cite{Narain1986NuclPB}, the momentum lattice structure is given by the Narain lattice
\lceq{
L = \Pi_{1,1} \oplus \Pi_{1,1} \oplus E_8(-1) \oplus E_8(-1) \, ,
}
while the symmetry enhancement appears when $p^2 = -2 $.\footnote{The full list of the allowed enhancements with the corresponding gauge algebras is worked out in \cite{Cvetic2022PRD, Font:2020rsk} with the help of the results of elliptic $K3$ fibrations \cite{shimada2001classification, shimada2005elliptic}.} The symmetry enhancement points are given by the rational quadratic divisor 
\lceq{
H(-1) = \bigcup_{(v,v) = -2, \, v \in L } v^\perp \, ,
}
 requiring that the lattice admits a $2$-reflective modular form (the lattice $L$ is $2$-reflective). This is the case for  the even unimodular lattice $\Pi_{2,18}$. 
 
 A weight $132$ modular form $\Psi_{(2,18)}(Z)$ can be obtained by applying the Borcherds product to the nearly holomorphic modular form \cite{Atsuhira2022} 
\lceq{\label{eq:sec8_264nhmf}
f(\tau) = \frac{E_4}{\Delta} (\tau)  = \frac{1}{q} + 264 + 8244 q + 139520 q^2  + \ldots \, ,
}
where $q = e^{2 \pi i \tau}$. $E_4, E_6$ are Eisenstein series with the constant term normalized to $1$   
\lceq{\label{eq:sec5_E4E6}
E_4 (\tau) = 1 + 240 \sum_{n=1}^\infty \frac{n^3 q^n}{1 - q^n} = 1+ 240 q + 2160 q^2 + \ldots  \, ,\\
E_6 (\tau) = 1 - 504 \sum_{n=1}^\infty \frac{n^5 q^n}{1 - q^n} = 1 - 504 q - 6632 q^2 + \ldots \, ,
}
and $\Delta(\tau) = \eta^{24} (\tau)$ is a weight $12$ cusp form where $\eta(\tau)$ is the Dedekind eta function  
$\eta(\tau) = q^{1/24} \prod_{n=1}^\infty (1-q^n)$. From Theorem~\ref{Thrm:sec4_RQD} we know that the modular form $\Psi_{(2,18)}(Z)$ is holomorphic (all the coefficient in the principal part is positive) and only has zeroes at the rational quadratic divisor $H(-1)$. Moreover, since the lattice $L$ is even unimodular now, the character for the group $\SO^+(L)$ must be trivial.

\begin{itemize}[leftmargin=*]
  \item \textbf{Signature $(2,10)$} 
\end{itemize}
For $l = 10$, if the level of the lattice is prime, there are ten types of reflective lattices. 

The simplest of these is the self-dual lattice $L = \Pi_{1,1} \oplus \Pi_{1,1} \oplus E_8$, which is 2-reflective. Requiring that the zeroes of $\Psi(Z)$ are contained in the rational quadratic divisors defined by $(v,v) = -2$, we should look for a nearly holomorphic modular form of weight $-4$ as an input into the Borcherds product. Such a function exists 
\lceq{
f(\tau) = \frac{E_4^2}{\Delta} (\tau)  = \frac{1}{q} + 504 + 73764 q + \ldots \, , \quad q = e^{2 \pi i \tau} \, ,
}
and the corresponding Borcherds product $\Psi(Z)$ is of weight $252$. Due to the unimodularity, the character for this lattice is, as required, trivial. Comparison of the possible gauge symmetry enhancements allowed by this lattice  would \cite{Hamada2021} would exclude this lattice.

Since all other lattices are at level $p >1$,  the enhancement points will correspond not only to short roots (vectors with norm $-2$) as for even unimodular lattices. 
Indeed, for reflective lattices roots are not only vectors with norm $-2$ but also vectors $v$ with norm $-2p$ satisfying $(v,u) = 0\mod 2$ for all vectors $u \in L$. In fact, the last condition is equivalent to saying that $v/p$ is in the dual lattice $L'$.

The most interesting class is for $p=2$. It contains three lattices, all of which have reflective vectors of norm $-2$ and $-4$ ($p=2$). The CHL lattice \cite{mikhailov1998} of the form
\lceq{\label{eq:latCHL}
L = \Pi_{1,1} \oplus \Pi_{1,1} (2) \oplus E_8(-1) \cong \Pi_{1,1} \oplus \Pi_{1,1} \oplus D_8(-1) \, 
}
is among these three. The full list of enhancements and the allowed gauge algebras in the $8D$ CHL theories \cite{Chaudhuri:1995fk, Chaudhuri:1995bf} is worked out in \cite{Font:2021uyw, Cvetic:2021sjm}, which shows that the symmetry enhancement points are exactly the orthogonal subspace defined by the reflective vectors of norm $-2$ and $-4$. In other words, the rational quadratic divisors of the modular form $\Psi_{(2,10)}(Z)$ should contain the following components 
\lceq{
H(-1) = \bigcup_{(v,v) = -2, \, v \in L } v^\perp \quad  \text{and} \quad \bigcup_{(v,v) = -4, \, v/2 \in L' } v^\perp \, .
}
The lattice satisfies the condition in Theorem~\ref{Thrm:sec4_trivial_character} thus the character of the modular group is trivial. Such reflective modular forms are closely related to many interesting study about Enriques surfaces  \cite{BORCHERDS1996Topology,gritsenko2010reflective, kondo2002JAG, gritsenko2015moduli}. We define the lattice $L_E = \Pi_{1,1} \oplus \Pi_{1,1} (2) \oplus E_8(-2)$. Notice that 
\lceq{
 L_E'(2) \cong \Pi_{1,1} \oplus \Pi_{1,1}(2) \oplus E_8(-1) \cong L \, ,
 }
 the orthogonal group has the following relation
 \lceq{
 \mO^+(L_E) \cong \mO^+(L_E') \cong \mO^+(L_E'(2)) \cong \mO^+(L) \, .
 }
Hence the (reflective) modular forms with respect to the group $\mO^+(L)$ correspond to that of the group $\mO^+(L_E)$. Note that the reflective vectors defined above are in the lattice $L$. For a modular form with respect to the group $\mO^+(L_E)$ (lattice $L_E$), we should check the relation of the reflective vectors and RQDs to those for the lattice $L$. Under the transformation $L_E \subset L_E' \rightarrow L_E'(2) \cong L$ (one can think that each vector is scaled by $\sqrt{2}$), the $2$ reflective vectors of $L_E$ and $4$ reflective vectors ($v_E \in L_E, \, v_E/2 \in L_E'$) transform to the $4$-reflective vectors and the $2$-reflective vectors of $L$ respectively. This correspondence can be summarized as follows:
\lceq{
\notag 
\Psi_1(Z): \, \text{4-reflective for lattice $L_E$} \longrightarrow  \text{2-reflective for lattice $L$} \, , \\
\Psi_2(Z): \, \text{2-reflective for lattice $L_E$} \longrightarrow \text{4-reflective for lattice $L$}   \, .
}
$\Psi_2(Z)$ is called the Borcherds-Enriques modular form $\Phi_4$, first found in \cite{BORCHERDS1996Topology} and reconstructed as an example in \cite{Borcherds1998} (see Example 13.7). The existence of the 4-reflective modular forms $\Psi_1(Z)$  for lattice $L_E$ is confirmed in \cite{gritsenko2015moduli} (Lemma 5.4). Therefore we can see that this case is indeed consistent. More explicitly, we can obtained the modular form $\Psi_{1,2} (Z)$ by lifting two vector-valued modular forms $F_{1,2} (\tau)$, which are obtained through the lifting of the $\Gamma_0 (2)$-modular form on $0$ respectively \cite{Scheithauer2017}
\lceq{
\label{eq:sec4_F1andF2}
 16\frac{\eta^8 (2 \tau)}{\eta^{16} (\tau)}, \quad  \frac{\eta^8(\tau)}{\eta^{16}(2 \tau )}.
}
The Borcherds lift of $F_1 (\tau) + F_2 (\tau)$ will generate the suitable modular form $\Psi_{(2,10)} (Z) = \Psi_1 (Z) \Psi_2 (Z)$ of weight $8$ that has the appropriate RQDs. Notice that we have the following identity 
\lceq{
\label{eq:sec4_F1plusF2}
 \frac{\eta^8 (\tau)}{\eta^{16} (2\tau)} + 2^4\frac{\eta^8 (2 \tau)}{\eta^{16} (\tau)} = \frac{E_4 (\tau)}{\eta^8(\tau) \eta^8(2 \tau)}, 
}
this is consistent with the sting-theoretic result. 

For the two other lattices in this class ($p=2$), $D_8$ factor of the CHL lattice in \eqref{eq:latCHL} is replaced by respectively $D_4 \oplus D_4$ and $D'_8$.\footnote{Notice that further reduction of the CHL strings to seven and six dimensions yields  $\Pi_{1,1} \oplus \Pi_{1,1} \oplus \Pi_{1,1} \oplus  D_4\oplus D_4(-1)$ and $\Pi_{1,1} \oplus \Pi_{1,1} \oplus \Pi_{1,1} \oplus \Pi_{1,1} \oplus D'_8(-2)$  lattices respectively \cite{mikhailov1998}.} The counterterms can again be obtained by a direct multiplication of two different modular forms.\footnote{For any other prime $p$ the lattices of signature $(2,10)$ do not admit modular forms that can be decomposed into a product of $2$-reflective and $2p$-reflective modular forms (Theorem 4.3 in \cite{wang2022}). Note that there are four $l=10$ lattices which are $2$-reflective and have non-prime level $p$ \cite{wang2022}.
Only the CHL lattice yields the gauge symmetry enhancement consistent with the swampland considerations.} In these cases, the $2$-reflective modular forms  are  of weight $60$ and $28$ respectively,  and the $4$-reflective modular forms are of weight $12$ and $28$ respectively. 

We conclude this section with several comments on the  proposed counterterms and the anomaly cancellation:
\begin{itemize}
    \item The functions that appear here only allow to recover the coefficients before the $\tr R^4$ term in the eight-dimensional quantum corrected effective action, as can also be checked by  comparing to the string theory computations. For example, in $l = 18$ Narain lattice case, the one-loop gravitational thresholds in the toroidally compactified heterotic string is \cite{Kiritsis:1997hf}
    \lceq{
    \mathcal{I}^{\text{het}}_{D} = -\mathcal{N} (2 \pi)^2 \int_{\SL(2;\Z) \backslash \mH} \frac{d^2 \tau}{\tau_2^2} (\tau_2)^{d/2}\Gamma_{d,d+16} \hat{\mathcal{A}}(\tau, R),
    }
    where $d = 10 - D$, $\tau_2 = \Im \tau$, $\Gamma_{d,d+16}$ is the lattice sum (the theta function corresponds to Narain lattice when $d = 2$) and the $\hat{A} (\tau, R)$ is
    \lceq{\label{eq:r4susy}
    \hat{\mathcal{A}}(\tau, R) =  \frac{1}{2^7 \cdot 3 ^2 \cdot 5} \frac{E_4 (\tau)}{\eta^{24} (\tau)} t_8 \tr R^4 + \frac{1}{2^9 \cdot 3^2} \frac{\hat{E}^2_2 (\tau)}{\eta^{24} (\tau)} t_8 (\tr R^2)^2 + \cdots . 
    } 
    The coefficient of the $\tr R^4$ part indeed agrees with our result. The functions appearing before $(\tr R^2)^2$ terms and further quartic involving gauge fields, denoted by ellipsis in \eqref{eq:r4susy}, are all different due to supersymmetry. This can be traced to the appearance of terms quadratic in curvature due to the nontrivial Bianchi identity for three-form $H$. These functions are related by recursion relations, that (from worldsheet perspective) can be found e.g. in \cite{Kiritsis:1997hf}. Extracting log parts in \eqref{eq:r4susy}, allows to recover tha anomaly polynomial \eqref{eq:anomalypol}. When decompactification  to ten dimensions is possible the complete counterterm must lift to ten-dimensional Green-Schwarz counterterm. These considerations do not affect our main argument on requirement that in eight-dimensions only reflective lattices are admissible.
    
    \item The construction of the counterterms based solely on the anomaly cancellation (and supergravity) allow some ambiguity.  Even after restricting  reflective lattices, there might be more than one reflective modular forms with appropriate RQDs exist and it is not yet clear how to uniquely fix the candidate counterterms without string theory arguments.\footnote{As an example, in $l=10$ case, $L_E$ admits another 2-reflective function different from~\eqref{eq:sec4_F1andF2}, which is of the weight $124$ (for more details, we refer readers to Lemma 5.4 in \cite{gritsenko2015moduli}). In principle it is also a good candidate for constructing the counterterm, though it seems not compatible with string theory.}
    \item So far we only deal with $l \ge 3$ cases. On the other hand, the $l = 2$ case suffers the same anomaly and requires similar counterterms. Though, stricktly speaking, Borcherds product does not apply to this case, one can try to construct the appropriate modular forms on group $\SO(2,2;\R)$ by taking advantage of the two to one group homomorphism from $\SL(2,\R) \times \SL(2, \R) $ to $\SO(2,2;\R)$ and consider the multiplications of two usual $\SL(2;\Z)$ (or its congruence subgroup) modular forms. We leave the detailed discussion to future work. 
\end{itemize} 
\section{Discussion}\label{sec:conclusion}
The moduli space of the eight-dimensional minimal supergravities  coupled to $l$ Yang-Mills multiplets is given by
\lceq{\notag
\mM = \frac{\SO(2,l)}{\U(1) \times \SO(l)} \, .
}
The composite $\U(1)$ connection, under which the fermions of the theory are chirally charged, is anomalous. The gauge fixing translates this anomaly into an anomaly under the discrete part of the coset denominator, which can be shown to coincide with the discrete modular group of the corresponding lattice. The consistency of the theory requires a suitable counterterm to cancel this discrete anomaly. 

The counterterms can be constructed with the use of the Borcherds product of the modular forms on the orthogonal group, $\Psi(Z)$:
\lceq{\label{eq:ccco}
\mathcal{S} = \frac{1}{r} \int \arg (\Psi(Z)) X_8 (R, \mF) \, ,
}
where $X_8(R, \mF)$ is the anomaly polynomial and $r$ is the weight of the modular form satisfying some conditions required by the anomaly cancellation. These conditions can be summarized as
\begin{itemize}
    \item The character for the modular group $\SG(L)$ (or the lattice $L$) must be trivial or the modular group $\SG (L)$ breaks into its subgroup such that the character is trivial for all the elements among this subgroup.
    \item The zeros and poles of $\Psi(Z)$ lie on the rational quadratic divisors. If these points can be interpreted as the symmetry enhancement points, it requires $\Psi(Z)$ should be reflective modular form and $L$ is the reflective lattice.
\end{itemize}
\vspace{0.2cm}
\noindent
We will conclude by outlining some open questions and directions for further research.

\vspace{0.2cm}
\noindent\textbf{Relation to the Swampland} $\quad$ 
It is not surprising that we find a larger set of theories with a mechanism for anomaly cancellation than what is allowed by swampland considerations. It is however curious, that there are finite number of admissible lattices  and they are bounded by $26$. In fact, the only two lattices  for $l>2$ that are believed to lead to consistent theories of quantum gravity  \cite{ParraDeFreitas2023} are even more special and admit $2$-reflective modular forms. It would be of great interest to find out if there exist physical requirements that lead to further constraints on the lattice structure.

Notice that we always assume the lattice to be even. This condition enters crucially in the construction of the modular forms on the orthogonal groups, and it is hard to see how a counterterm can be constructed otherwise. We do not know a more direct supergravity (swampland?) argument in support of this condition that arises very naturally in string theory.

\vspace{0.2cm}
\noindent\textbf{Counterterms and massive sectors} $\quad$ In our $\NN=1$ discussion the precise form of the anomaly polynomial played no role. In fact \eqref{eq:sec3_anomalous_phase} is computed only by knowing the massless spectrum. On the other hand, the string amplitudes receive contributions from massive states. For a very recent interesting discussion of importance of these see \cite{Montero:2022vva}.
At the supergravity level one could generate corrections to the counterterm to \eqref{eq:sec3_anomalous_phase} by adding massive states and integrating them out. It is hard to believe that the choices of massive sector are arbitrary, and as discussed in \cite{Minasian2017} one expects that reduction on $\mathbb{P}^1$ to six-dimensional $(1,0)$ would impose strong constraints on the possible massive sectors. The question of whether and when a theory admits different consistent massive completions is certainly of great interest. 

\vspace{0.2cm}
\noindent\textbf{$K3$ reductions and 4D physics} $\quad$ It is also of interest to explore the implications of the 8D counterterms discussed here for compactifications, particularly 4D couplings. There are very direct parallels between 8D  minimally supersymmetric theories and 4D $\NN=2$ theories. 

The $K3$ reduction of 8D theory with 16 supercharges to a 4D $\NN=2$ theory parallels the reduction of 10D heterotic strings on $K3$. There, a separate integration of the Bianchi identity (with the constraints that the instanton numbers should sum up to 24) and of the Green-Schwarz term yield two different four-forms that agree with those obtained in the factorised anomaly polynomial in the resulting 6D $(1,0)$ theory (see e.g. \cite{Duff:1996rs}). So one could wonder about similar reduction of the counterterm in 8D.

Choosing an instanton in group $H \subset G$ (rank$(G)=l$) breaks the gauge group to $G_0$ stabilised by $H$ in $G$. The Bianchi identity can be written in general as (following the notation of \cite{Hamada2021}) 
$$
d H_3 = \kappa \tr R^2 + \ell \cdot \tr \mF ^2
$$
where $\kappa$ can take values 1 or 0 (only for $l=2$), and $\ell$ is the level of the current algebra (for a product gauge group, summation over different gauge factors is implied), and hence  $\ell \cdot c_2(H) =  24 \kappa$.\footnote{ For $\kappa = 0 $, there cannot be nontrivial gauge configurations over $K3$. The reduction yields a 4D $\mathcal{N}=2$ theory with three vector multiplets and 20 neutral hypermultiplets.}
Denoting rank$(H)=h$, 
$$
\SO(2,l) \, \longrightarrow \ \SO(2, l-h) \, .
$$
But in 4D, $n_V = l-h+1$, and the extra multiplet comprises one of the vectors in 8D gravity multiplet, and the dilaton-axion. Notice that while in 8D the counterterm must have nontrivial modular properties, the 4D threshold corrections $\sim \tr R^2$ involve automorphic functions on $\SO(2,n_V)$. The addition of the extra scalar (``conformal compensator'' in vector moduli space) should be responsible for this change. It would be of some interest to understand how this works in more detail.

It has been argued that the $K3$ reduction of $\NN=1$ theories in 8D provides a good framework for studying 4D $\NN=2$ compactifications since it encompasses not only the $K3\times T^2$ but also the heterotic flux backgrounds \cite{Melnikov:2012cv}. Considering the space of all 8D $\SO(2,l)$ for $l=2,10,18$ would enlarge this space and hopefully cover all $\NN=2$ theories of heterotic type, i.e. those for which the dilaton is in the vector multiplets. This raises an interesting possibility that all threshold corrections in these theories would in some way be governed and be derivable from the special  $\SO(2,l)$ modular forms from which the counterterms \eqref{eq:ccco} are built.

\section*{Acknowledgements} We thank Peng Cheng, Jonathan Heckman, Renata Kallosh, Ilarion Melnikov, Nikita Nekrasov, Boris Pioline, Valentin Reys, Raffaele Savelli, Yi Shan, Stefan Theisen and Yu-Xiao Xie for useful communications and conversations. Special thanks are due to Guillaume Bossard, Hector Parra De Freitas, Jim Liu and Haowu Wang. The work of RM is partially supported by ERC grants 772408-Stringlandscape and 787320-QBH Structure.

\appendix
\section{Orthogonal modular forms}
\label{app:supplement_modular_form}

Some necessary properties of orthogonal modular forms were reviewed in  subsection \ref{sub:orthogonal_modular_forms}. 
 In order to make the paper more self-contained, more background material is collected in this Appendix. Definitions and theorems are given without proofs. Our presentation follows closely  \cite{bruinier2002borcherds},  which can be consulted for detailed explanations. 

Throughout this appendix, as in the main text, we denote by $L$ an even lattice of signature $(b^+, b^-) = (2,l)$ and assume $l \ge 3$.


\subsection{The Weil representation}
We denote the complex upper-half plane $\mH = \{\tau \in \C ; \, \Im \tau > 0\}$.  $\tau$ is the standard variable on $\mH$ and  we use $x$ and $y$ for its real and imaginary parts respectively ($\tau = x + i y$). For $z \in \C$ we define $e(z) =e^{2 \pi i z}$ and denote by $\sqrt{z} = z^{1/2}$ the principal branch of the square root. For arbitrary $b \in \C$, we define $z^{b} = e^{b \Ln z}$ where $\Ln z$ denotes the principal branch of the logarithm. We denote by $\Mp (2; \R) $  the metapletic group, i.e. the double cover of group $\SL(2;\R)$, realized by the two choices of holomorphic square roots of $\tau \rightarrow c \tau + d $ for arbitrary element $ M \in \SL(2;\R)$,
\lceq{
M = \mx{
a & b \\
c & d
}, \quad a,b,c,d \in \R, \quad \det M = ad - bc = 1 \, .
}
Any element in $\Mp(2;\R)$ can be written as $\left(M, \phi(\tau)\right)$ where $M \in \SL(2,\R)$ and $\phi(\tau)^2 = c \tau + d$. The multiplication in the group $\Mp (2;\R)$ is defined as 
\lceq{
\left(M_1 , \phi_1 (\tau )\right) \left(M_2, \phi_2(\tau )\right) = \left(M_1 M_2, \phi_1 (M_2 \tau ) \phi_2 (\tau )\right) \, ,
}
where $M \tau  = (a \tau  + b)/(c \tau + d)$ denotes the usual action of $\SL(2;\R)$. By fixing the choice  $\phi(\tau) = \sqrt{ c \tau + d}$, we actually define a locally isomorphic embedding $\SL(2;\R) \xhookrightarrow{} \Mp(2;\R) $
\lceq{
M  \mapsto \widetilde{M} = \left(M, \sqrt{c \tau + d}\right). 
}
$\Mp(2;\Z)$ is generated by two elements $T, S$ 
\lceq{
T = \left(\mx{
1 & 1 \\
0 & 1
}, 1\right) \, , \quad  S = \left(\mx{
0 & -1 \\
1 & 0
}, \sqrt{\tau }\right) \, .
}
One has the relation $S^2 = (ST)^3 = Z $, where 
\lceq{
Z = \left(\mx{
-1 & 0 \\
0  & -1
}, i\right)
}
is the standard generator of the center of $\Mp(2;\Z)$. For convenience we define $\Gamma_1 = \SL(2;\Z)$, 
\lceq{
\Gamma_\infty= \left\{
\mx{
1 & n\\
0 & 1
}; \, n\in \Z  
\right\} \le \Gamma_1 \, , \\
\widetilde{\Gamma}_{\infty} = \langle T \rangle  = \left\{
\left(
\mx{
1 & n\\
0 & 1
} , 1
\right) ; \, n\in \Z
\right\} \, ,
}
where $\langle T \rangle $ denotes the group generated by $T$. 

Suppose $L$ is an even lattice equipped with a symmetric $\Z$-valued bilinear form $\left(z_1,z_2\right)$ for $z_1,z_2 \in L$ and the associated quadratic form $q (z) = (z,z)/2$ is integer for arbitrary $z \in L$. We denote by $L'$ the dual lattice. The quotient $L' / L$ is a finite Abelian group, the so-called discriminant group. Since the quadratic form can be extended to the dual lattice, we can define the quadratic form on $L'/L$, which takes values in $\Q/\Z$. There is a unitary representation $\varrho$ of $\Mp(2;\Z)$ on the algebra $\C [L'/L]$. If we denote the standard basis of $\C [L'/L]$ by $\left\{\mathfrak{e}_\gamma | \gamma \in L'/L \right\}$, then $\varrho$ can be defined by the action of the generators $S,T \in \Mp(2;\Z)$ as follows
\longeq{
& \varrho (T) \mfe_\gamma = e(q(\gamma)) \, ,\\
& \varrho (S) \mfe_\gamma = \frac{\sqrt{i}^{b^-  - b^+}}{\sqrt{|L'/L |}} \sum_{\delta \in L'/L} e(-(\gamma, \delta)) \mfe_\delta \, .
}
This is the so-called Weil representation. Based on the relation $S^2 = Z $, we have 
\longeq{ \label{eq:sec4_actionZ}
\varrho (Z) \mfe_\gamma & =  \frac{i^{b^- - b^+}}{|L'/L |} \sum_{\delta, \lambda \in L'/L} e(-(\gamma, \delta )) e(-(\delta, \lambda )) \mfe_\lambda \\
& = i^{b^-  - b^+ } \mfe_{-\gamma} \, .
}
We denote by  $\langle \cdot, \cdot \rangle $ the standard product of $\C [L' / L]$, i.e.
\lceq{
\left\langle \sum_{\gamma \in L'/L} \lambda_\gamma \mfe_\gamma , \sum_{\gamma \in L'/L} \mu_\gamma \mfe_\gamma  \right\rangle = \sum_{\gamma  \in L'/L} \lambda_\gamma \bar{\mu}_\gamma \, .
}
For $\gamma,\delta \in L'/L$, we can define the representation matrix element $\varrho_{\gamma \delta } (M,\phi) = \left\langle \varrho(M, \phi) \mfe_\delta , \mfe_\gamma \right\rangle $.

\subsection{Vector-valued modular forms}
\begin{defn}\label{defn:sec4_petersson_slash}
(Petersson slash operator) Let $\kappa \in \frac{1}{2} \Z$ and $f$ be a $\C[L'/L]$-valued function on $\mH$. For $(M,\phi) \in \Mp(2;\Z)$ we define the Petersson slash operator $|_\kappa (M,\phi)$ by
\lceq{
\left(f |_\kappa (M,\phi)\right) (\tau) = \phi(\tau)^{-2 \kappa} \varrho(M,\phi)^{-1} f(M\tau ) \, .
}
\end{defn}
We denote by $\varrho^*$ the dual representation of $\varrho$. If we think of $\varrho(M,\phi)$ as a matrix with entries in $\C$, then $\varrho^*(M,\phi)$ is simply the complex conjugate of $\varrho(M,\phi)$. The "dual operation" of $\Mp(2;\Z)$ on functions $f: \mH \rightarrow \C[L'/L]$ is given by 
\lceq{
\left(f|^*_\kappa (M,\phi)\right) (\tau) = \phi(\tau)^{-2 \kappa } \varrho^* (M,\phi)^{-1} f(M \tau) \, .
}
If we assume that the function $f : \mH \rightarrow \C [L'/L]$ is a  holomorphic function which is invariant under the $|_\kappa^*$ operation of $T \in \Mp(2;\Z)$. Since $f$ can be expanded by the basis $\mfe_\gamma$ of $L'/L$, we have $f = \sum_{\gamma } f_\gamma  \mfe \gamma$. The invariance is satisfied if and only if 
\lceq{\label{eq:sec4_periodic_f}
f_\gamma (\tau) = f_\gamma|_\kappa^* T (\tau ) = e^*(q (\gamma))^{-1} f_\gamma (\tau +1) \\ \Leftrightarrow \quad e(q (\gamma) \tau) f_\gamma(\tau ) = e\left(q(\gamma)(\tau+1)\right) f_\gamma(\tau+1) \, ,
}
which means the invariance of $f$ under $T$ implies that the function $e(q(\gamma) \tau) f_\gamma (\tau)$ is periodic with period $1$. We can directly Fourier expand $f$ by 
\lceq{
f(\tau) e(q(\gamma) \tau)  = \sum_{\gamma \in L'/L} \sum_{n \in \Z } c(\gamma,n ) e(n \tau) \mfe_\gamma \, .
}
To have a compact expression, we define $\mfe_\gamma (n \tau) = e(n \tau) \mfe_\gamma$ and write
\lceq{
f(\tau) = \sum_{\gamma \in L'/L} \sum_{n \in \Z - q(\gamma)} c(\gamma,n) \mfe_\gamma(n \tau) \, ,
}
with Fourier coefficients 
\lceq{
c(\gamma, n) = \int_0^1 \langle f (\tau), \mfe_\gamma( n \bar{\tau}) \rangle d x \, .
}
\begin{defn}\label{defn:appB_hmf_dual_Weil}
(holomorphic modular form of dual Weil representation) Let $\kappa \in \frac{1}{2} \Z$. A function $f: \mH \rightarrow \C [L]$ is called a modular form of weight $\kappa$ with respect to $\varrho^*$ and $\Mp(2;\Z)$ if
\begin{enumerate}[label=\roman*)]
  \item $f|_\kappa^* (M, \phi) = f$ for all $(M,\phi) \in \Mp(2;\Z)$,
  \item $f$ is holomorphic on $\mH$, 
  \item $f$ is holomorphic at the cusp $\infty$. If $c(\gamma, 0) \equiv 0$, $f$ is called a cusp form.
\end{enumerate}
\end{defn}
The condition $(\rom{3})$ requires $f$ has a Fourier expansion of the form
\lceq{
f(\tau ) = \sum_{\gamma \in L'/L} \sum_{ \substack{n \in \Z - q(\gamma ) \\
n \ge 0} } c(\gamma, n) \mfe_\gamma (n \tau) \, .
}
The $\C$-vector space of modular forms of weight $\kappa$ with respect to $\varrho^*$ and $\Mp(2;\Z)$ is denoted by $M_{\kappa, L}$ and the subspace of cusp forms  is denoted by $S_{\kappa,L}$ . Similar to the usual complex valued modular form of $\SL(2;\Z)$, the linear space $M_{\kappa,L}$ is finite dimensional.

\subsection{Nearly holomorphic modular forms}
\begin{defn}\label{defn:appB_nhmf} 
(nearly holomorphic modular form) A function $f: \mH \rightarrow \C[L]$  is called a nearly holomorphic modular form of weight $k$ (with respect to $\varrho$ and $\Mp(2;\Z)$), if 
\begin{enumerate}[label=\roman*)]
    \item $f|_k (M, \phi) = f$ for all $(M,\phi) \in \Mp(2;\Z)$, 
    \item $f$ is holomorphic on $\mH$,
    \item $f$ has a pole in $\infty$, i.e. $f$ has a Fourier expansion of the form
    \lceq{
    f(\tau) = \sum_{\gamma \in L'/L } \sum_{ \substack{n \in \Z + q(\gamma) \\ n \gg - \infty}} c(\gamma, n ) \mfe_{\gamma} (n \tau) \, .
    }
\end{enumerate}
The space of these nearly holomorphic modular forms is denoted by $M_{k,L}^!$. The summation $n \gg - \infty$ indicates that there exists a finite negative number $n_0$ such that all $n \ge n_0$. This condition implies that the pole at the cusp ($\infty$) has finite order. The Fourier polynomial 
\lceq{
\sum_{\gamma \in L'/L} \sum_{\substack{n \in \Z + q(\gamma ) \\ n <0}} c(\gamma, n) \mfe_\gamma (n \tau) 
}
is called the principal part of $f$. 
\end{defn}
As shown in \cite{bruinier2002borcherds}, the space of nearly holomorphic modular form is generated by the Poincar\'{e} series, thus is finite dimensional. The principal part should satisfy the Theorem \ref{Thrm:appB_serre_dual}. 

\subsection{Modular forms on generalized upper-half plane}
The orthogonal modular forms and Borcherds product were introduced in the main text. Recall the definition of $j(M, Z)$ in (\ref{eq:sec4_action_M4})).  More generally we can rewrite it as
\lceq{
j(M, Z) = \left(M Z_L, z\right) \, ,
}
where $z = \left(1,0,\ldots, 0\right)^T$ is the $l+2$ vector and $Z_L = (- q_0 (Z), Z, 1)^T$. Suppose $L$ is an even lattice of signature $(2,l)$ and $V = L \otimes \R$. The function $j(M,Z)$ on $\mO^+ (V) \times \mH_l$ is an automorphy factor for $\mO^+(V)$, i.e. it satisfies the cocycle relation
\lceq{
j(M_1 M_2 ,Z) = j(M_1, M_2 \langle Z \rangle) j (M_2, Z) \, .
}
For an arbitrary $a \in \C$, we have already specified $\Arg a \in [-\pi, \pi)$, which is the principal value of argument of $a$. We denote by $\Ln$ the logarithm of the principal branch, which is defined as $\Ln a = \ln |a| + i \Arg a$. For an arbitrary $a,b \in \C$, we define $a^b = e^{b \, \Ln a}$. Let $r \in \Q$, if $M \in \mO^+ (V)$ and $Z \in \mH_l$, then $j(M,Z)^r = e^{r \Ln j(M,Z)}$. There exists a map $w_r$ from $\mO^+(V) \times \mO^+(V)$ to the set of roots of unity (of order bounded by the denominator of $r$) such that
\lceq{
j(M_1 M_2, Z)^r = w_r (M_1, M_2) j(M_1,M_2 \langle Z\rangle)^r j(M_2, Z)^r \, .
}
\begin{defn}\label{defn:sec3_multiplier_system}
(multiplier system) Let $\Gamma \le \mO^+(V)$ be a subgroup and $r \in \Q$ as above. By a multiplier system of weight $r$ we mean a map 
\lceq{
\chi: \Gamma \longrightarrow S^1 = \left\{t \in \C| \, |t| = 1\right\}
}
satisfying 
\lceq{
\chi (M_1 M_2) = w_r (M_1,M_2) \chi(M_1) \chi(M_2) \, , \quad M_1, M_2 \in \Gamma \, .
}
If $r \in \Z$, then $\chi$ is actually a character of $\Gamma$, then $\chi(M) j(M,Z)^r$ is a cocycle of $\Gamma$. 
\end{defn}
\begin{defn}\label{defn:sec4_mf_on_guhp}
(modular form on generalized upper-half plane) Let $\Gamma \le \Gamma(L)$ be a subgroup of finite index and $\chi$ a multiplier system for $\Gamma$ of weight $r \in \Q$. A meromorphic function $F$ on $\mH_l$ is called a meromorphic modular from of weight $r$ and multiplier system $\chi$ with respect to $\Gamma$, if 
\lceq{ \label{eq:sec4_modular_form}
\Psi(M \langle Z \rangle ) = \chi (M) j(M, Z)^r  \Psi(Z)
}
for all $M \in \Gamma$. If $\Psi$ is even holomorphic on $\mH_l$ then it is called a holomorphic modular form. 
\end{defn}
The Borcherds product can lift a nearly holomorphic modular form $f(\tau) = \sum_{\gamma \in L'/L} f_\gamma \mfe_\gamma: \mH \rightarrow \C[L'/L]$ (see Definition~\ref{defn:appB_nhmf}) of weight $1 - l/2$ with Fourier expansion
\lceq{
f(\tau) = \sum_{\gamma \in L'/L} \sum_{n \in \Z + q(\gamma)} c(\gamma,n) \mfe_\gamma(n \tau) \, ,
}
to the meromorphic function $\Psi(Z) : \mH \rightarrow \C$ of weight $c(0,0)/2$. The precise theorem is stated as follows.
\begin{thrm}\label{Thrm:sec4_Borcherds}
(Theorem 13.3 (1) in \cite{Borcherds1998} or Theorem 3.22 (\rom{1}) in \cite{bruinier2002borcherds}) Let $L$ be an even lattice of signature $(2,l)$ with $l \ge 3$, and $z \in L$ a primitive isotropic vector. Let $z' \in L'$ and $K = L \cap z^\perp \cap {z'}^\perp$. Moreover, assume that $K$ also contains an isotropic vector. Let $f$ be a nearly holomorphic modular form of weight $k = 1 - l/2$ whose Fourier coefficients $c(\gamma, n )$ are integral for $n < 0$. Then
\lceq{
\Psi (Z)  = \prod_{\beta \in L'/L} \prod_{\substack{m \in \Z + q(\beta) \\  m<0}} \Psi_{\beta, m}(Z)^{c(\beta,m)/2}
}
is a meromorphic function on $\mH_l$ of (rational) weight $c(0,0)/2$ for the modular group $\Gamma(L)$ with some multiplier systems $\chi$ of finite order. If $c(0,0) \in 2 \Z$, then $\chi $ is the character of group $\Gamma(L)$.  
\end{thrm}
For functions $\Psi_{\beta ,m }(Z)$ see Definition 3.14 in \cite{bruinier2002borcherds}. 

We can turn to the zeros and poles of $\Psi(Z)$. A nowhere-vanishing holomorphic modular forms $\Psi(Z)$ obtained through Borcherds product cannot exist since there is no input nonzero holomorphic modular form $f(\tau)$ of negative weight $1 - l/2$. Before determining the positions of poles and zeroes, it is necessary to explain the concept of rational quadratic divisors (Heegner divisors).

Let $z \in L$ be a primitive norm $0$ vector, $z' \in L'$ with $(z, z') = 1$. Let $N$ be unique positive integer with such that $(z, L) = N \Z$. Then we have $z/N \in L'$. Denote by $K$ the lattice
\lceq{
K = L \cap z^{\perp} \cap z'^{\perp} \, .
}
 $K$ has signature $(b^+ - 1, b^- -1 ) = (1, l -1 )$. For an arbitrary vector $n \in V = L \otimes \R$,  $n_K$ denotes  the orthogonal projection $n$ to $K \otimes \R$ and
\lceq{\label{eq:appC_projectionK}
n_K = n - (n,z) z' + (n, z) (z', z') z - (n, z') z \, .
}
If $n \in L'$, then $n_K$ lies in the dual lattice $K'$ of $K$.  Let $\zeta \in L$  be a lattice vector with $(\zeta, z) =N$. Let $n \in L$, then the vector 
\lceq{
\tilde{n} = n - (n,z/N) \zeta-(n,z')z + (n,z/N)(\zeta,z')z
}
lies in $L$ and easy to verify that $\tilde{n} \perp z$ and $\tilde{n} \perp z'$. Hence $\tilde{n} \in K$ and each element $n \in L$ can be uniquely decomposed in this way, or equivalently, $L = K \oplus \Z \zeta \oplus \Z z$. Now let $\lambda \in L'$  be a vector of negative norm, i.e. $q (\lambda) < 0$. Then the orthogonal complement $\lambda^\perp \subset L \otimes \R$ is a rational quadratic space of type $(2,l-1)$. With these settings we can define the rational quadratic divisors
\begin{defn}\label{defn:sec3_RQD}
(rational quadratic divisor or Heegner divisor) Let $\lambda \in L'$ be a vector of negative norm $m$, we set 
\lceq{
H_\lambda = \left\{[Z_L] \in \mathcal{K}^+| \, (Z_L, \lambda) = 0\right\} \, .
}
Moreover, due to the decomposition $Z_L = (-q(Z) - q(z')) z + Z + z'$ (recall the equation~(\ref{eq:sec4_decomposeZ})) and $\lambda =  b z + \lambda_K + a z' $, expanding the inner product $(Z_L, \lambda)$ yields 
\lceq{
H_\lambda \cong \left\{Z \in \mH_l | \, a q(Z)  - (Z, \lambda_K) - a q (z') - b = 0\right\} \, 
}
in coordinates on $\mH_l$. This set defines a prime divisor on $\mH_l$. Suppose $\beta \in L' / L$ and $m$ is a negative rational number; the sum 
\lceq{
H(\beta, m) = \sum_{\substack{\lambda \in \beta + L \\ q(\lambda) = m}} H_\lambda 
}
is called the rational quadratic divisor (or Heegner divisor) of discriminant $(\beta, m )$, which is a $\Gamma(L)$-invariant divisor on $\mH_l$. When $\beta = 0$, we usually denote $H(m) = \frac{1}{2}H(0,m)$.
\end{defn}
This definition is suitable for lattices of signature $(2,l)$ with arbitrary Gram matrix. If we specify the Gram matrix of $L = \Pi_{1,1} \oplus L_0$ as defined in the equation~(\ref{eq:sec4_qudratic_form}) and the vector $z, z'$, equivalently we have
\lceq{
H_\lambda = \left\{Z \in \mH_l| \, a q_0 (Z) - (Z, \lambda_K)_0 - b = 0\right\} \, ,
}
where the subscript emphasizes that the inner product is associated with the quadratic form $S_0$. With this definition we can describe the position of the zeros and poles by the following theorem. 
\begin{thrm}\label{Thrm:sec4_RQD}
(Theorem 13.3 (2) in \cite{Borcherds1998} or Theorem 3.22 (\rom{2}) in \cite{bruinier2002borcherds})
The zeros and poles of $\Psi(Z)$ lies on the divisor of $\Psi(Z)$ on $\mH_l$, which is the linear combinations of Heegner divisors determined by the principal part of the nearly holomorphic modular form $f$
\lceq{
    \left(\Psi\right) = \frac{1}{2} \sum_{\beta \in L'/L} \sum_{\substack{m \in \Z + q(\beta) \\ m<0}} c(\beta, m) H(\beta, m) \, .
}
The multiplicities of $H(\beta, m)$ are $2$, if $2\beta = 0$ in $L'/L$, and $1$, if $2\beta \neq 0$ in $L'/L$.
\end{thrm}
As we saw from the above theorems, the properties of the Borcherds product $\Psi(Z)$ are completely captured by the nearly holomorphic modular form $f(\tau)$, in particular by the principal part of $f(\tau)$:
\lceq{
\sum_{\gamma \in L'/L} \sum_{\substack{n \in \Z + q(\gamma) \\ n < 0}} c(\gamma,n) \mfe_\gamma(n \tau) \, .
}
Pairing the form $f(\tau)$ with a vector valued cusp form of weight $1 + l /2$ for the dual Weil representation (see Definition~\ref{defn:appB_hmf_dual_Weil}) gives a meromorphic elliptic modular form of weight $2$ for $\SL(2;\Z)$, hence its constant term must vanish by the residue theorem (no nonzero $\SL(2;\Z)$ modular form of weight $2$) and this gives the conditions on the principal part on $f$, stated as the following theorem. By setting $\kappa = 1 + l /2$ and denoting the space of the vector valued modular cusp form of weight $\kappa$ with respect to lattice $L$ as $S_{\kappa,L}$, we have
\begin{thrm}\label{Thrm:appB_serre_dual}
(Theorem 1.17 in \cite{bruinier2002borcherds}) There exists a nearly holomorphic modular form $f \in M_{k,L}^!$ with prescribed principal part 
\lceq{
\sum_{\beta \in L'/L} \sum_{\substack{m \in \Z +q (\beta) \\ m<0}} c(\beta, m) \mfe_\beta (m \tau) 
}
($c(\beta, m ) \in \C$ with $c(\beta, m) = c(- \beta, m)$), if and only if the functional 
\lceq{
\sum_{\beta \in L'/L}  \sum_{\substack{m \in \Z +q (\beta) \\ m<0}} c(\beta, m) a_{\beta, -m} ,
}
equals zero in $S_{\kappa,L}^*$. For $\gamma \in D(L)$ and $n \in \Z - q(\gamma)$ with $n > 0$, $a_{\gamma,n}: S_{\kappa, L} \rightarrow \C$ denote the functional in the dual space $S_{\kappa,L}^*$ of $S_{\kappa,L}$ which maps a cusp form $f$ to its $(\gamma,n)$-th Fourier coefficient $a_{\gamma ,n } (f)$. 
\end{thrm}
Obviously this imposes non-trivial condition on the principal part of the nearly holomorphic modular form $f$. 

\subsection{Character of the lattice}
If the weight of the modular form is integer, which is the  case of interest, the multiplier system is actually the character of the modular group $\Gamma(L)$, or the character of the lattice $L$.  This forms a homomorphism from the modular group to $\U(1)$. From the well-known Pontryagin duality, the abelianisation $G^{ab} = G/[G,G]$ of the group $G$ is isomorphic to the character group $\text{Hom}(G,\C^\times)$. Thus to obtain the character we need to consider the abelianisation of the modular group $\Gamma(L)$. As discussed in section~\ref{sec:general_strategy}, the anomaly cancellation imposes the triviality of the character for the admissible lattices. To the best of our knowledge,  the sufficient and necessary conditions for a lattice of signature $(2,l)$ to have trivial characters are not known. A sufficient condition is known:

\begin{thrm}\label{Thrm:sec4_trivial_character}
(Theorem 1.7 in \cite{GRITSENKO2009JA}) Let $L$ be an even integral lattice containing at least two hyperbolic planes ($\Pi_{1,1}$), such that $\text{rank}_3(L) \ge 5$,\footnote{For any prime $p$ the $p$-rank of L, denoted by $\text{rank}_p(L)$, is the maximal rank of the sublattices $M$ such that $\det(M)$ is coprime to $p$.} and $\text{rank}_2(L) \ge 6$, then the $\Gamma(L)^{ab} \cong \Z/2\Z$ and $\SG(L)^{ab}$ is trivial, where $\SG(L)$ is the modular group intersect with the special orthogonal group of lattice $L$, i.e. $\SG(L):=\Gamma(L) \cap \SO(L)$. 
\end{thrm}
An immediate corollary is that if $L = \Pi_{1,1} \oplus \Pi_{1,1} \oplus \hat{L}$ and $\hat{L}$ contains a sublattice isomorphic to $A_2$, it satisfies the so-called Kneser conditions~\cite{Kneser1984, GRITSENKO2009JA} and the character for group $S\Gamma(L)$ is trivial. Notably, if the lattice $L = \Pi_{1,1} \oplus \Pi_{1,1} \oplus \hat{L}$ is an even unimodular lattice of rank at least $6$, we have the same conclusion that the $\Gamma(L)^{ab} \cong \Z/2\Z$ and $\SG(L)^{ab}$ is trivial.


\bibliography{Refs}



\end{document}